\documentclass[a4paper]{aa}
\usepackage{graphics,txfonts,longtable}
\usepackage{rotating}

\def\teff{$T_{\rm eff}$}
\def\lgg{$\log g$}

\def\vs{$v_{\rm e}\sin i$}
\def\vmacro{$V_{\rm macro}$}
\def\gef{$g_{\rm eff}$}

\newcommand{\ha}{H$\alpha$}

\newcommand{\bs}{$\langle B_{\rm s}\rangle$}
\newcommand{\kms}{km\,s$^{-1}$}
\newcommand{\ms}{m\,s$^{-1}$}

\newcommand{\firps}[1]{\resizebox{\hsize}{!}{\rotatebox{-90}{\includegraphics{#1}}}}

\newcommand{\firrps}[2]{\resizebox{#1}{!}{\rotatebox{-90}{\includegraphics{#2}}}}

\def\i{\,{\sc i}} \def\ii{\,{\sc ii}} \def\iii{\,{\sc iii}}

\begin{document}

\title{Pulsation tomography of rapidly oscillating Ap stars%
\thanks{Based on observations made with the SAO 6-m telescope, with the Canada-France-Hawaii Telescope, and the ESO VLT
(DDT programme 274.D-5011 and programme 072.D-0138, retrieved
through the ESO archive).}} \subtitle{Resolving the third dimension
in peculiar pulsating stellar atmospheres}

\titlerunning{Pulsation tomography of rapidly oscillating Ap stars}
\authorrunning{T. Ryabchikova et al.}

\author{T.~Ryabchikova\inst{1,2}
\and
M.~Sachkov\inst{2}
\and
O.~Kochukhov\inst{3}
\and
D.~Lyashko\inst{4}
}

\offprints{T. Ryabchikova, \\ \email{ryabchik@inasan.ru}}

\institute{Department of Astronomy, University of Vienna,
T\"urkenschanzstrasse 17, A-1180 Wien, Austria
\and
Institute of Astronomy, Russian Academy of Sciences, Pyatnitskaya 48, 119017 Moscow, Russia
\and
Department of Astronomy and Space Physics, Uppsala University Box 515, SE-751 20 Uppsala, Sweden
\and
Tavrian National University, Yaltinskaya 4, 330000 Simferopol, Ukraine}

\date{Received / Accepted }

\abstract {} {We present detailed analysis of the vertical pulsation
mode cross-section in ten rapidly oscillating Ap (roAp) stars based
on spectroscopic time-series observations. 
The aim of this analysis
is to derive from observations a complete picture of how the
amplitude and phase of magnetoacoustic waves depend on depth. } 
{We use the unique properties of roAp stars, in particular chemical
stratification, to resolve the vertical structure of $p$-modes. Our
approach consists of characterising pulsational behaviour of a
carefully chosen, but extensive sample of spectral lines. We analyse
the resulting amplitude-phase diagrams and interpret observations in
terms of pulsation wave propagation.} {We find common features in
the pulsational behaviour of roAp stars. Within a sample of
representative elements the lowest amplitudes are detected for
Eu\ii\ (and Fe in 33 Lib and in HD~19918), then pulsations go
through the layers where H$\alpha$ core, Nd, and Pr lines are
formed. There RV amplitude reaches its maximum, and after that
decreases in most stars. The maximum RV of the second REE ions is
always delayed relative to the first ions. The largest phase shifts
are detected in Tb\iii\ and Th\iii\ lines. Pulsational variability
of the Th\iii\ lines is detected here for the first time. The Y\ii\
lines deviate from this picture, showing even lower amplitudes than
Eu\ii\ lines but half a period phase shift relative to other weakly
pulsating lines. We measured an extra broadening, equivalent to a
macroturbulent velocity from 4 to 11--12 \kms\ (where maximum values
are observed for Tb\iii\ and Th\iii\ lines), for pulsating REE
lines. The surface magnetic field strength is derived for the first
time for three roAp stars: HD~9289 (2 kG), HD~12932 (1.7 kG), and
HD~19918 (1.6 kG). } { The roAp stars exhibit similarity in the
depth-dependence of pulsation phase and amplitude, indicating
similar chemical stratification and comparable vertical mode
cross-sections. In general, pulsations waves are represented by a
superposition of the running and standing wave components. In the
atmospheres of roAp stars with the pulsation frequency below the
acoustic cut-off frequency, pulsations have a standing-wave
character in the deeper layers and behave like a running wave in the
outer layers. Cooler roAp stars develop a running wave higher in the
atmosphere. In stars with pulsation frequency close to the acoustic cut-off one, 
pulsation waves have a running character starting from deep
layers. The transition from standing to running wave is accompanied
by an increase in the turbulent broadening of spectral lines. }


\maketitle

\section{Introduction}
\label{intro}

About 10\,\% to 20\,\% of upper main sequence stars are
characterised by remarkably rich line spectra, often containing
numerous unidentified features. Compared to the solar case,
overabundances of up to a few dex are often inferred for some
iron-peak and rare-earth elements, whereas some other chemical
elements are found to be underabundant (Ryabchikova et al.
\cite{RNW04}). Some of these \textit{chemically peculiar} (Ap) stars
also exhibit organised magnetic fields with a typical strength of a
few kG. Chemical peculiarities are believed to result from the
influence of the magnetic field on the diffusing ions, possibly in
combination with the influence of a weak, magnetically-directed wind
(e.g., Babel \cite{Babel92}).

More than 30 cool Ap stars exhibit high-overtone, low-degree, non-radial $p$-mode pulsations
with periods in the range of 6--21 minutes (Kurtz \& Martinez \cite{KM00}), with their
observed pulsation amplitudes modulated according to the visible magnetic field structure.
These \textit{rapidly oscillating Ap} (roAp) stars are key objects for asteroseismology, which
presently is the most powerful tool for testing theories of stellar structure and evolution.

Recent progress in observational studies of roAp stars has been
achieved by considering high time-resolution spectroscopy in
addition to the classical high-speed photometric measurements.
High-quality time-resolved measurements of magnetic pulsators have
revealed a surprising diversity in the pulsational behaviour of
different lines in the roAp spectra (e.g., Kanaan \& Hatzes
\cite{KH98}; Kurtz et al. \cite{KEM06} and references therein).
Detailed analyses of the bright roAp star $\gamma$~Equ (Savanov et
al. \cite{SMR99}; Kochukhov \& Ryabchikova \cite{KR01a}) demonstrate
that spectroscopic pulsational variability is dominated by the lines
of rare-earth ions, especially those of Pr and Nd, which are strong
and numerous in the roAp spectra. On the other hand, light and
iron-peak elements do not pulsate with amplitudes above 50--100~\ms,
which is at least an order of magnitude lower in comparison with the
1--5~\kms\ variability observed in the lines of rare-earth elements
(REE). Many other roAp stars were  proven to show very similar
overall pulsational behaviour (e.g., Kochukhov \& Ryabchikova
\cite{KR01b}; Balona \cite{Balona02}; Mkrtichian et al.
\cite{MHK03}; Kurtz et al. \cite{KEM05a}), with an exceptional case
in 33~Lib, which shows small-amplitude variations in Fe lines, and
possibly two other stars, HD~12932 and HD~19918, where RV variations
in the Fe\i\ 5434.52~\AA\ line have been marginally detected (Kurtz
et al. \cite{KEM05b}).

The early spectroscopic studies speculated that the unusual
diversity of the pulsation signatures in roAp spectra can be
attributed to an interplay between the $p$-mode pulsation geometry
and inhomogeneous horizontal or vertical element distributions (see
discussions in Kanaan \& Hatzes \cite{KH98}; Savanov et al.
\cite{SMR99}; Baldry \& Bedding \cite{BB00}; Mkrtichian et al.
\cite{MHP00}). However, none of these studies present models capable
of explaining pulsations in real stars. Peculiar characteristics of
the $p$-mode pulsations in a roAp star were finally clarified by
Ryabchikova et al. (\cite{RPK02}), who were the first to empirically
determine vertical stratification  of chemical elements and relate
chemical profiles to pulsational variability. In their study of the
atmospheric properties of $\gamma$~Equ, Ryabchikova et al. show that
the light and iron-peak elements are enhanced in the lower
atmospheric layers, whereas REE ions are concentrated in a cloud
with a lower boundary at $\log\tau_{5000}\la-4$ (Mashonkina et al.
\cite{MRR05}). Thus, high-amplitude pulsations observed in REE lines
occur in the upper atmosphere, while lines of elements showing no
significant variability form in the lower atmosphere. This leads to
the following general picture of roAp pulsations: we observe a
signature of a magnetoacoustic wave, propagating outwards through
the chemically stratified atmosphere with increasing amplitude.

In addition to remarkable pulsational behaviour, the REE lines
formed in the upper atmospheric layers of roAp stars exhibit an
extra broadening, corresponding to a macroturbulent velocity
\vmacro\,=\,10~\kms, which cannot be attributed to the chemical
stratification or magnetic field effects (Kochukhov \& Ryabchikova
\cite{KR01a}). In the recent detailed line-profile variability
study, Kochukhov et al. (\cite{KR07}) have presented evidence for
the existence of peculiar asymmetric oscillation patterns in the
broad REE lines of several roAp stars. It was demonstrated that the
inferred pulsation signatures can be reproduced with the spectrum
synthesis calculations, which contain an extra pulsational line
width variability in addition to the usual velocity perturbations.
These results suggest that a turbulence zone modulated by pulsations
probably exists in the upper atmospheres of roAp stars.

The presence of significant phase shifts between pulsation radial
velocity (RV) curves of different REEs (Kochukhov \& Ryabchikova
\cite{KR01a}) or even lines of the same element (Mkrtichian et al.
\cite{MHK03}) can be attributed to the chemical stratification
effects and, possibly, to a short vertical length of running
magnetoacoustic wave. However, it is not immediately clear whether
all spectroscopic observations of roAp stars can be fitted into this
simple picture and to what extent magnetoacoustic pulsation theories
can explain these observations. The wide diversity of pulsation
signatures (in particular, phases and bisector variability) of the
REE lines probing similar atmospheric heights and, especially, the
presence of pulsation node in the atmosphere of 33~Lib (Mkrtichian
et al. \cite{MHK03}) are inexplicable in the framework of the
non-adiabatic pulsation models developed by Saio \& Gautschy
(\cite{SG04}) and Saio (\cite{S05}). These calculations correctly
predict an increase in pulsation amplitude with height, but show
neither nodes nor rapid phase changes in the REE line-forming
region. To resolve this discrepancy it is therefore imperative to
analyse in detail pulsational variations of many different ions in
the spectra of representative sample of roAp stars. Only in this way
can we derive meaningful observational constraints for pulsation
theories and search for regular patterns in the pulsation
characteristics of different roAp stars.

First general results for roAp stars were presented by Kurtz at al.
(\cite{KEM05a}). They considered bisector behaviour of the H$\alpha$
core and the Nd\iii\,6145~\AA\ line in 10 roAp stars. Based on the
H$\alpha$ core measurements, the authors point out that the
increasing amplitude with height in the atmosphere is a common
characteristic of all stars in their sample. Kurtz et al.
(\cite{KEM05a}) also note that in some stars, e.g. HD~12932,
pulsations have a standing-wave behaviour, while more complex
pulsations are observed in some other stars. These results were
obtained from the analysis of bisector variability of one particular
line, Nd\iii\,6145~\AA. This feature is not optimal for the bisector
analysis because of the different blending effects both in the line
core and the line wings that depend on the effective temperature
and/or chemical anomalies.

In this study we have embarked on the task of obtaining a detailed
vertical cross-section of the roAp pulsation modes. We used the
unique properties of roAp stars, in particular chemical
stratification, to resolve the vertical structure of $p$-modes and
to study propagation of pulsation waves at a level of detail, that
was previously only possible for the Sun. Our \textit{pulsation
tomography} approach consists of characterising the pulsational
behaviour of a carefully chosen, but extensive, sample of spectral
lines including weak ones and subsequently interpretating the
observations in terms of pulsation wave propagation. Therefore, our
sample was limited to slowly rotating roAp stars. The aim of this
analysis is to derive observationally a complete picture of how the
amplitude and phase of magnetoacoustic waves depend on depth and
to correlate resulting vertical mode cross-sections with other
pulsational characteristics and with the fundamental stellar
parameters. Furthermore, we envisage that unique 3-D maps of roAp
atmospheres and non-radial pulsations can be derived by combining
pulsation tomography results presented here with the horizontal
pulsation maps obtained with Doppler imaging (Kochukhov \cite{K04})
or moment analysis (Kochukhov \cite{K05}).

Our paper is organised as follows. Section~\ref{observ} describes
acquisition and reduction of roAp time-series spectra. The choice of
targets and physical properties of the stellar sample are discussed
in Sect.~\ref{parameters}. Radial velocity measurements and period
analysis are presented in Sect.~\ref{RV}. Pulsation tomography
results are summarised in Sect.~\ref{ph-a}. Bisector variability is
analysed in Sect.~\ref{bis}. Results of our study are summarised and
discussed in Sect.~\ref{disc}.

\section{Observations and data reduction}
\label{observ}

\begin{table*}[!th]
\caption{Log of time-series observations for roAp stars.}
\label{Table_UVES_Log}
\begin{center}
\begin{tabular}{lccccccl}
\hline
\hline
~~~Star    &  Start HJD & End HJD  & Number of &  Exposure & Overhead & Peak & ~Telescope/Instrument \\
           &  (2450000+)&(2450000+)& exposures &  time (s) & time (s) & $S/N$   & ~~~(observing mode) \\
\hline
HD\,9289  & 2920.54506 & 2920.62881 & 111 & 40 & 25 & 90  & VLT/UVES (600~nm) \\
HD\,12932 & 2921.62234 & 2921.70532 &  69 & 80 & 25 & 90  & VLT/UVES (600~nm) \\
HD\,19918 & 2921.52607 & 2921.60905 &  69 & 80 & 25 & 100 & VLT/UVES (600~nm) \\
HD\,24712 & 3321.65732 & 3321.74421 &  92 & 50 & 22 & 300 & VLT/UVES (390+580~nm)    \\
HD\,101065& 3071.67758 & 3071.76032 & 111 & 40 & 25 & 180 & VLT/UVES (600~nm) \\
HD\,122970& 3069.70977 & 3069.79359 & 111 & 40 & 25 & 160 & VLT/UVES (600~nm) \\
HD\,128898& 3073.80059 & 3073.88262 & 265 & 1.5& 25 & 250 & VLT/UVES (600~nm) \\
HD\,134214& 3070.77571 & 3070.85848 & 111 & 40 & 25 & 260 & VLT/UVES (600~nm) \\
HD\,137949& 3071.76312 & 3071.84598 & 111 & 40 & 25 & 350 & VLT/UVES (600~nm) \\
HD\,201601& 2871.46470 & 2871.56295 &  70 & 80 & 42 & 80  & SAO 6-m/NES (425--600~nm) \\
HD\,201601& 2186.70618 & 2186.80296 &  64 & 90 & 43 & 250 & Gecko/CFHT(654--6660~nm) \\
HD\,201601& 2186.82456 & 2186.92308 &  65 & 90 & 43 & 230 & Gecko/CFHT(662--673~nm) \\

\hline
\end{tabular}
\end{center}
\end{table*}

The main observational dataset analysed in our study consists of 958
observations of 8 roAp stars, obtained with the UVES instrument at
the ESO VLT between October 8, 2003 and March 12, 2004 in the
context of the observing programme 072.D-0138 (Kurtz et al.
\cite{KEM06}). The ESO Archive facility was used to search and
retrieve science exposures and the respective calibration frames.
Observations of each target covered 2 hours and consisted of an
uninterrupted high-resolution spectroscopic time-series with a total
number of exposures ranging from 69 to 265. The length of individual
exposures was 40$^{\rm s}$ or 80$^{\rm s}$, except for the brightest
roAp star HD\,128898 ($\alpha$~Cir), for which a 1.5$^{\rm s}$
exposure time was used. The ultra-fast (625kHz/4pt) readout mode of
the UVES CCDs allowed us to limit overhead to $\approx$\,$20^{\rm
s}$, thus giving a duty cycle of 70--80\% for the majority of the
targets. The signal-to-noise ratio (S/N) of individual spectra is
between 90 and 350, as estimated from the dispersion of the stellar
fluxes in the line-free regions. Detailed description of
observations for each target is presented in
Table~\ref{Table_UVES_Log}. Columns of the table give the stellar
name, heliocentric Julian dates for the beginning and the end of
spectroscopic monitoring, number of observations, exposure and
overhead times, peak signal-to-noise ratio of individual spectra,
and information about the telescope and instrument where the data
were obtained.

The red arm of the UVES spectrometer was configured to observe the
spectral region 4960--6990~\AA\ (central wavelength 6000~\AA). The
wavelength coverage is complete, except for a 100~\AA\ gap centred
at 6000~\AA. Observations were obtained with the high-resolution
UVES image slicer (slicer No. 3), providing an improved radial
velocity stability and giving maximum resolving power for
$\lambda/\Delta\lambda\approx115\,000$.

All spectra were reduced and normalised to the continuum level with
a routine especially developed by one of us (DL) for a fast
reduction of spectroscopic time-series observations. A special
modification of the Vienna automatic pipeline for \'echelle spectra
processing (Tsymbal et al. \cite{tsymbal}) was developed. All bias
and flat field images were median-averaged before calibration. The
scattered light was subtracted by using a 2-D background
approximation. For cleaning cosmic ray hits, we used an algorithm
that compares the direct and reversed spectral profiles. To
determine the boundaries of \'echelle orders, the code used a
special template for each order position in each row across
dispersion axes. The shift of the row spectra relative to template
was derived by a cross-correlation technique. Wavelength calibration
was based on a single ThAr exposure, recorded immediately after each
stellar time series. Calibration was done by the usual 2-D
approximation of the dispersion surface. An internal accuracy of
30--40~\ms\ was achieved by using several hundred ThAr lines in all
\'echelle orders. The final step of continuum normalisation and
merging of the \'echelle order was carried out by transformation of
the flat field blaze function to the response function in each
order.

The global continuum normalisation was improved by iteratively
fitting a smoothing spline function to the high points in the
average spectrum of each roAp star. With this procedure we corrected
an underestimate of the continuum level, unavoidable in analysis of
small spectral regions of the crowded spectra of cool Ap stars.
Correct determination of the absolute continuum level is important
for retrieving unbiased amplitudes of radial velocity variability
when the centre-of-gravity method is used. In addition to the global
continuum correction, spectroscopic time series were post-processed
to ensure homogeneity in the continuum normalization of individual
spectra. Extracted spectra were divided by the mean, the resulting
ratio was heavily smoothed, and then it was used to correct
continua in individual spectra. Without this correction a spurious
amplitude modulation of pulsation in variable spectral lines may
arise due to an inconsistent continuum normalisation.

The red 600~nm UVES dataset was complemented by the observations of
HD\,24712 obtained on November 11, 2004 in the DDT program
274.D-5011. Ninety-two time-resolved spectra were acquired with the
UVES spectrometer, configured to use the 390+580~nm dichroic mode
(wavelength coverage 3300--4420 and 4790--6750~\AA). A detailed
description of the acquisition and reduction of these data is given
by Ryabchikova et al. (\cite{RSW06}).

For the roAp star HD\,201601 ($\gamma$~Equ), we analysed 70 spectra
obtained on August 19, 2003 with the NES spectrograph attached to
the 6-m telescope of the Russian Special Astrophysical Observatory.
These \'echelle spectra cover the region 4250--6000~\AA\ and have
typical $S/N$ of $\approx$\,$80$. The data were recorded by
Kochukhov et al. (\cite{KRP04}), who searched for rapid magnetic
field variability in $\gamma$~Equ. We refer readers to that paper
for the details on the acquisition and reduction of the time-series
observations at the SAO 6-m telescope.

Post-processing of the \'echelle spectra of HD\,24712 and HD\,201601
was done consistently with the procedure adopted for the main
dataset. In addition, the time-resolved observations of HD\,201601
obtained in 2001 using the single-order $f/4$ Gecko coud\'e
spectrograph with the EEV1 CCD at the 3.6-m Canada-France-Hawaii
telescope were used. Observations have a resolving power of about
115\,000, determined from the widths of a number of ThAr comparison
lines. The reduction is described in Kochukhov et al. (\cite{KR07}).

\section{Fundamental parameters of programme stars}
\label{parameters}

\begin{table*}
\caption{Fundamental parameters of target stars.\label{tbl2}}
\begin{tabular}{rlccrcll}
\hline
\hline
HD~~~  & ~~Other         & \teff     & \lgg & \vs~~     & \bs & P          & Reference  \\
number & ~~name          & (K)       &      & (\kms)    & (kG)& (min)      &            \\
\hline
  9289 &BW Cet           & 7840      & 4.15 & 10.5~~    & 2.0 &{\it 10.522}& this paper \\
 12932 &BN Cet           & 7620      & 4.15 & 3.5~~     & 1.7 &{\it 11.633} & this paper \\
 19918 &BT Hyi           & 8110      & 4.34 & 3.0~~     & 1.6 &{\it 11.052} & this paper \\
 24712 &DO Eri, HR 1217  & 7250      & 4.30 & 5.6~~     & 3.1 &~{\it 6.125}, 6.282 & Ryabchikova et al. (\cite{RLG97}) \\  
101065 &Przybylski's star& 6600      & 4.20 & 2.0~~     & 2.3 &{\it 12.171} & Cowley et al. (\cite{CRK00}) \\        
122970 &PP Vir           & 6930      & 4.10 & 4.5~~     & 2.3 &{\it 11.187} & Ryabchikova et al. (\cite{RSH00}) \\  
128898 &$\alpha$~Cir     & 7900      & 4.20 & 12.5~~    & 1.5 &~{\it 6.802}, 7.34 & Kupka et al. (\cite{KRW96}) \\        
134214 &HI Lib           & 7315      & 4.45 & 2.0~~     & 3.1 &~{\it 5.690}& this paper \\                        
137949 &33 Lib           & 7550      & 4.30 &$\le$2.0~~~& 5.0 &~{\it 8.271},4.136, 9.422& Ryabchikova et al. (\cite{RNW04}) \\  
201601 &$\gamma$~Equ     & 7700      & 4.20 &$\le$1.0~~~& 4.1 &{\it 12.20~} & Ryabchikova et al. (\cite{RPK02}) \\  
\hline
\end{tabular}
\end{table*}

Fundamental parameters of the programme stars are given in
Table~\ref{tbl2}. For six stars, effective temperatures \teff,
surface gravity  \lgg, and mean surface magnetic fields \bs\ were
taken from the literature. For 4 remaining stars, HD\,9289,
HD\,12932, HD\,19918, and HD\,134214, atmospheric parameters were
derived using Str\"omgren photometric indices (Hauck \& Mermilliod
\cite{HM98}) with the calibrations by Moon \& Dworetsky
(\cite{MD85}) and by Napiwotzki et al. (\cite{N93}) as implemented
in the {\tt TEMPLOGG} code (Rogers \cite{R95}). In addition Geneva
photometric indices (Burki et al. \cite{B05})\footnote{{\tt
http://obswww.unige.ch/gcpd/ph13.html}} with the calibration of
K\"unzli et al. (\cite{KNK97}) was used for effective temperature
determination. The colour excesses were estimated from the reddening
maps by Lucke (\cite{L78}). In Table~\ref{tbl2} we present average
values of the effective temperatures derived with three different
calibrations. A typical dispersion is $\pm$150 K.

In all stars but HD\,101065, rotational velocities were estimated by
fitting line profiles of the magnetically insensitive Fe\i\ 5434.5,
5576.1 \AA\ lines. Magnetic spectral synthesis code {\tt SYNTHMAG}
(Piskunov \cite{P99}; Kochukhov \cite{synthmag06}) was used in our
calculations. Atomic parameters of spectral lines were extracted
from  the {\tt VALD} (Kupka et al. \cite{vald299}) and  {\tt DREAM}
(Bi\'emont et al. \cite{dream99}) databases, supplemented with the
new oscillator strengths for La\ii\, (Lawler et al. \cite{La2}),
Nd\ii\, (Den Hartog et al. \cite{Nd2-03}), Nd\iii\, (Ryabchikova et
al. \cite{RRKB06}), Sm\ii\, (Lawler et al. \cite{Sm2}), and Gd\ii\,
(Den Hartog et al. \cite{Gd2}). We confirmed rotational velocities
derived previously for HD\,24712, HD\,122970, HD\,128898,
HD\,137949, and HD\,201601. High spectral resolution of the present
data allows us to improve the value of the projected rotational
velocity for HD\,101065 (Przybylski's star) using partially resolved
Zeeman patterns in numerous lines of the rare-earth elements. The
value of the magnetic field modulus, 2.3 kG (Cowley et al.
\cite{CRK00}), was confirmed.

We note that the derived values of \vs\ in very sharp-lined roAp
stars depend strongly on the Fe stratification, which is known for
some of them (e.g. $\gamma$~Equ, see Ryabchikova et al.
\cite{RPK02}) and is typically characterised by the Fe overabundance
below $\log\tau_{5000}=-1$ and Fe depletion in the outer layers.
Rotational velocities in Table~\ref{tbl2} are derived for a
homogeneous Fe distribution. However, in all programme stars we see
observational evidence of Fe stratification, such as anomalous
strength of the high-excitation lines compared to the low-excitation
lines (see Ryabchikova et al. \cite{RWL03}). We tested one of the
most peculiar stars, 33~Lib, for possible influence of
stratification on the derived \vs\ and found that including
stratification may decrease the inferred rotational velocity to
1.5~\kms\ in comparison to 2.5--3.0~\kms\ obtained with a
homogeneous Fe abundance distribution. The lower value of \vs\ seems
to fit the observed resolved and partially resolved Zeeman
components better.

In three programme stars, HD\,9289, HD\,12932, and HD\,19918, mean
magnetic modulus was estimated for the first time from differential
magnetic broadening/intensification. Two spectral regions with a
pair of lines having different magnetic sensitivity were synthesised
for a set of magnetic field strengths. The first region contain
well-known Fe\ii\, 6147.7~\AA\ (\gef=0.83) and 6149.3~\AA\
(\gef=1.35) lines, while Fe\i\, 6335.3~\AA\ (\gef=1.16) and
6336.8~\AA\ (\gef=2.00) lines were analysed in the second spectral
region. The derived values of the magnetic fields were confirmed by
fitting other magnetically sensitive lines, for example, Fe\ii\,
6432.7~\AA\ (\gef=1.82) and Eu\ii\, 6437.6~\AA\ (\gef=1.76).

\section{Radial velocity measurements}
\label{RV}

To perform a meaningful study of the pulsational amplitudes in
spectral lines of different chemical elements/ions, one has to be
very careful in the choice of lines for pulsation measurements. For
this purpose we have synthesised the observed spectral region for
each star with the model atmosphere parameters and magnetic field
values from Table~\ref{tbl2}. Abundances for HD~24712, HD~101065,
HD~122970, HD~128898, HD~137949, and HD~201601 were taken from the
papers cited in the last column of Table~\ref{tbl2}. For the
remaining four stars, a preliminary abundance estimate was obtained
in this paper.

The radial velocities were measured with a centre-of-gravity
technique. We used only unblended or minimally blended lines. In
some cases where the line of interest was partially overlapping with
the nearby lines, only the unblended central part of the line was
considered; therefore, some lines were not measured between the
continuum points. This usually leads to lower pulsation amplitudes
if we have strong variations in the pulsation signal across a
spectral line (see $\gamma$~Equ -- Sachkov et al. \cite{sach04}, and
HD~99563 -- Elkin et al. \cite{EKM05}). Bisector radial velocity
measurements were performed for H$\alpha$ core and for a subset of Y
\ii, Eu\ii, Nd\ii, Nd\iii, Pr\ii, Pr\iii, Tb\iii, and Th\iii\
spectral lines. The pulsational RV variability of the two strongest
Th\iii\ lines at $\lambda\lambda$ 5376.13 and 6599.48~\AA\ was
investigated for several stars in our sample for the first time.

A detailed frequency analysis of the RV data for 8 stars from our
sample was carried out by Kurtz et al. (\cite{KEM06}). These authors
used the same observations as we do and, despite the relatively
short time span (2 hours) of the spectroscopic time series, they
claimed to resolve several frequencies for each star and to find
amplitude modulation that was not observed in photometry. Although a
detailed frequency analysis is not the primary goal of our paper, we
did repeat time-series analysis for all lines. First we applied the
standard discrete Fourier transformation (DFT) to the RV data. The
period corresponding to the highest pulsation amplitude value was
then improved by the sine-wave least-square fitting of the RV data
with pulsation period, amplitude, and phase treated as free
parameters. This fit was removed from the data and then Fourier
analysis was applied to the residuals. This procedure was repeated
for all frequencies with the S/N above 5.

To verify our analysis against the results of Kurtz et al.
(\cite{KEM06}), we applied it to the set of Pr\iii\, lines (to each
line separately and to the average Pr\iii\, RV) and compared in
detail the resulting solution for HD~134214. Our frequency solution
agrees perfectly with Kurtz et al. (\cite{KEM06}), although our RV
amplitudes are systematically lower. We believe that the primary
reason for this discrepancy, which is present for other stars as
well, is our systematically higher continuum placement, leading to
lower pulsation amplitude when the centre-of-gravity method is used.
As explained in Sect.~\ref{observ}, we have rectified observations
with a spline-fit over wide wavelength regions, whereas Kurtz et al.
(\cite{KEM06}) have probably assigned continuum points to the high
spectrum points in the immediate vicinity of the line considered.

\begin{figure*}[!th]
\centering
\firrps{82mm}{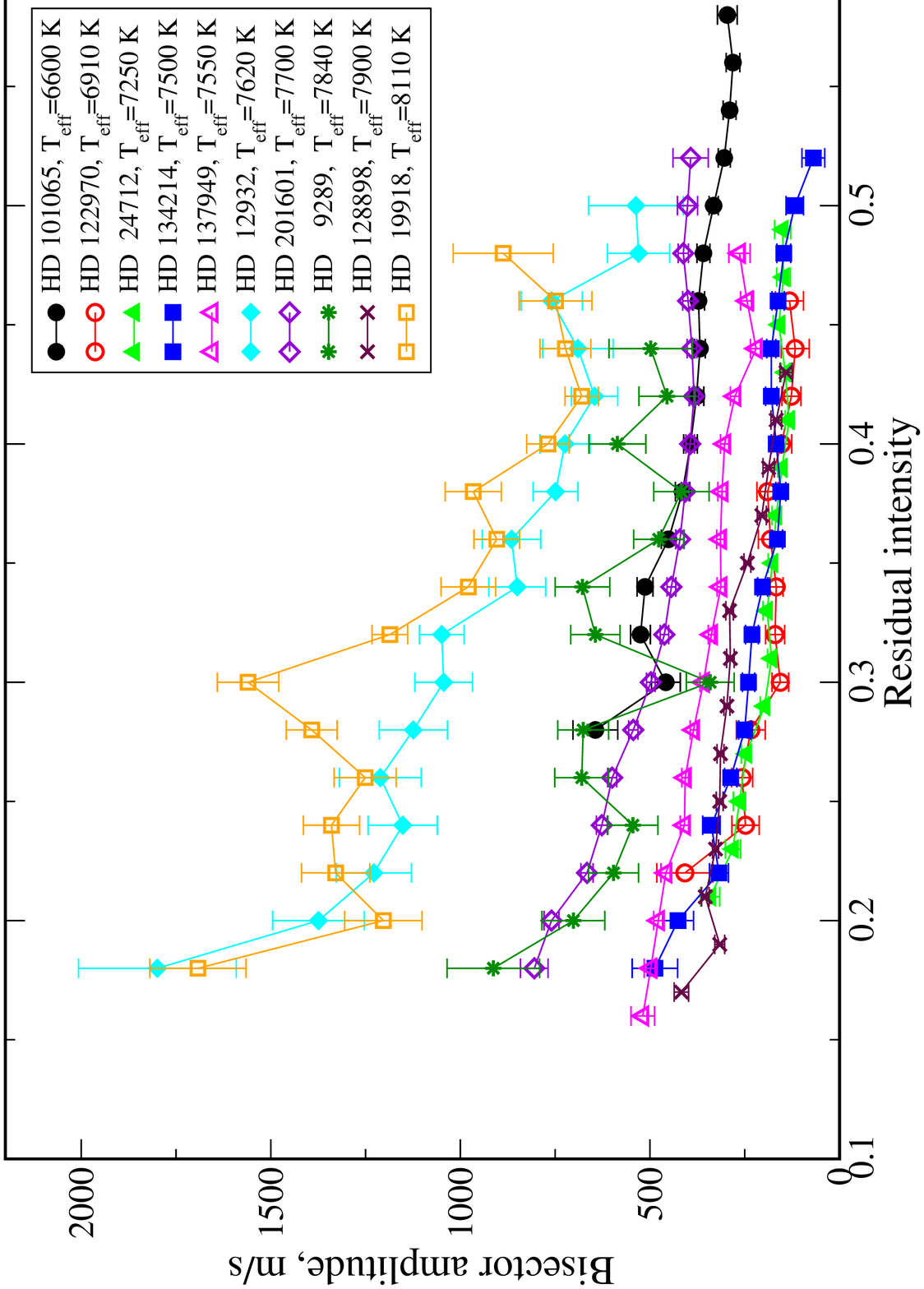}\hspace{0.5cm}\firrps{82mm}{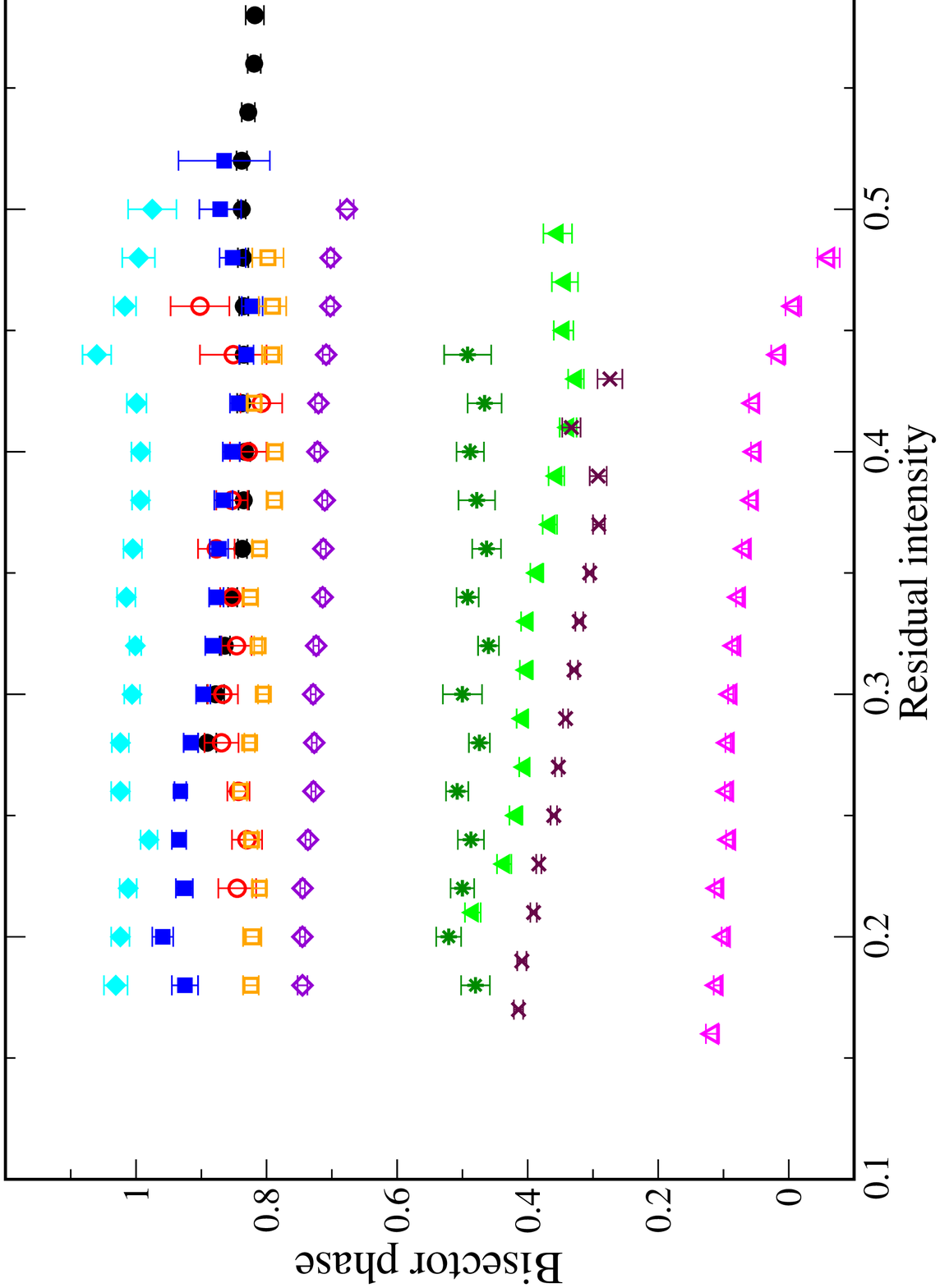}
\caption{Bisector measurements in the \ha\ core for the programme
stars. The RV amplitudes are shown in the left panel and pulsation
phases (based on the main periods from Table~\ref{tbl2}) in the
right panel.} \label{ha}
\end{figure*}

For each spectral line, we estimated the probability that the
detected periodicity is not due to noise (Horne \& Baliunas
\cite{HB86}). Then, for the lines with probabilities higher than
0.999, we calculated a weighted average value of the pulsation
period. All amplitudes and phases were then recalculated keeping the
period fixed. This information is summarised in Table~\ref{tbl3}
(available online only) for all lines studied in each star.
Table~\ref{tbl3} contains amplitudes (first line) and phases (second
line), together with the corresponding errors for the dominant
pulsation period. When no pulsation signal was detected, we give the
formal amplitude solution for the fixed period without phase
information. If a star had one dominant period, the RV analysis was
done with this period. In the cases of more than one dominant
period, a simultaneous fit was performed with up to three periods.
RV variation was approximated with the expression
\begin{equation}
\langle V \rangle = V_0 (t-t_0) + \sum_{i=1}^3 V_{i} \cos{\{2 \pi [(t-t_0)/P_{i} - \varphi_{i}]\}}.
\label{cosfit}
\end{equation}
Here the first term takes possible drift of the spectrograph's zero
point  into account. For all stars but HD~24712, HJD of the first
exposure of the star at a given night was chosen as a reference time
$t_0$. For HD~24712, HJD=2453320.0 was used as a reference time.
Both $V_{i}$ and $\varphi_{i}$ are, respectively, amplitude and
phase of the RV variability with the $i$th period ($i_{\rm max}=3$).
With the minus sign in front of $\varphi_{i}$, a larger phase
corresponds to a later time of RV maximum. This phase agreement is
natural when discussing effects of the outward propagation of
pulsation waves in the atmospheres of roAp stars.

The periods employed for determining of the RV amplitudes and phases
are given in the seventh column of Table~\ref{tbl2}. When more than
one period was inferred in the fitting procedure, the main period
for which pulsation amplitudes and phases are given in
Table~\ref{tbl3} (online material) is highlighted with italics.

\section{Bisector measurements}
\label{bis}

Radial velocity analysis was complemented by studying bisector
variability. Figure~\ref{ha} shows bisector amplitudes (left panel)
and phases (right panel) across the core of \ha\ line calculated
with the pulsation periods from Table~\ref{tbl2}. Where two periods
or the main period and its first harmonic are resolved, simultaneous
fit with two frequencies was done. In all programme stars, an
increase in bisector amplitude by two or more times from the
transition region to the deepest part of the core is observed. Thus
we supported the conclusion made by Kurtz et al. (\cite{KEM05a}).
This change is gradual in all but two stars: HD~9289 and HD~19918,
which have the lowest S/N. It is, therefore, unclear if small jumps
of RV amplitude across the core are real or caused by the low S/N of
the spectroscopic data. RV changes are accompanied by tiny phase
changes. Only in four stars do phase changes exceed the error bars.
These are the stars with the shortest pulsation periods close to the
acoustic cut-off frequencies: HD~24712, HD~128898, HD~132214, and
HD~137949 (33 Lib). The NLTE calculations show that, in the
atmosphere of the star with \teff\ between 7000 and 8000 K, the \ha\
core is formed at $-5\la\log\tau_{5000}\la-2$ (Mashonkina, private
communication). It is applied to a normal atmosphere; however, a
core-wing anomaly is present in all programme stars (Cowley~et~al.
\cite{CWA}), which was attributed to a peculiar atmospheric
stratification containing a region of increased temperature below
$\log\tau_{5000}=-4$ (Kochukhov~et~al. \cite{cwaT}). This change in
the atmospheric structure may lead to an upward shift in formation
depth of the base of the H$\alpha$ core (Mashonkina, private
communication).

\begin{figure*}[!th]
\centering
\firps{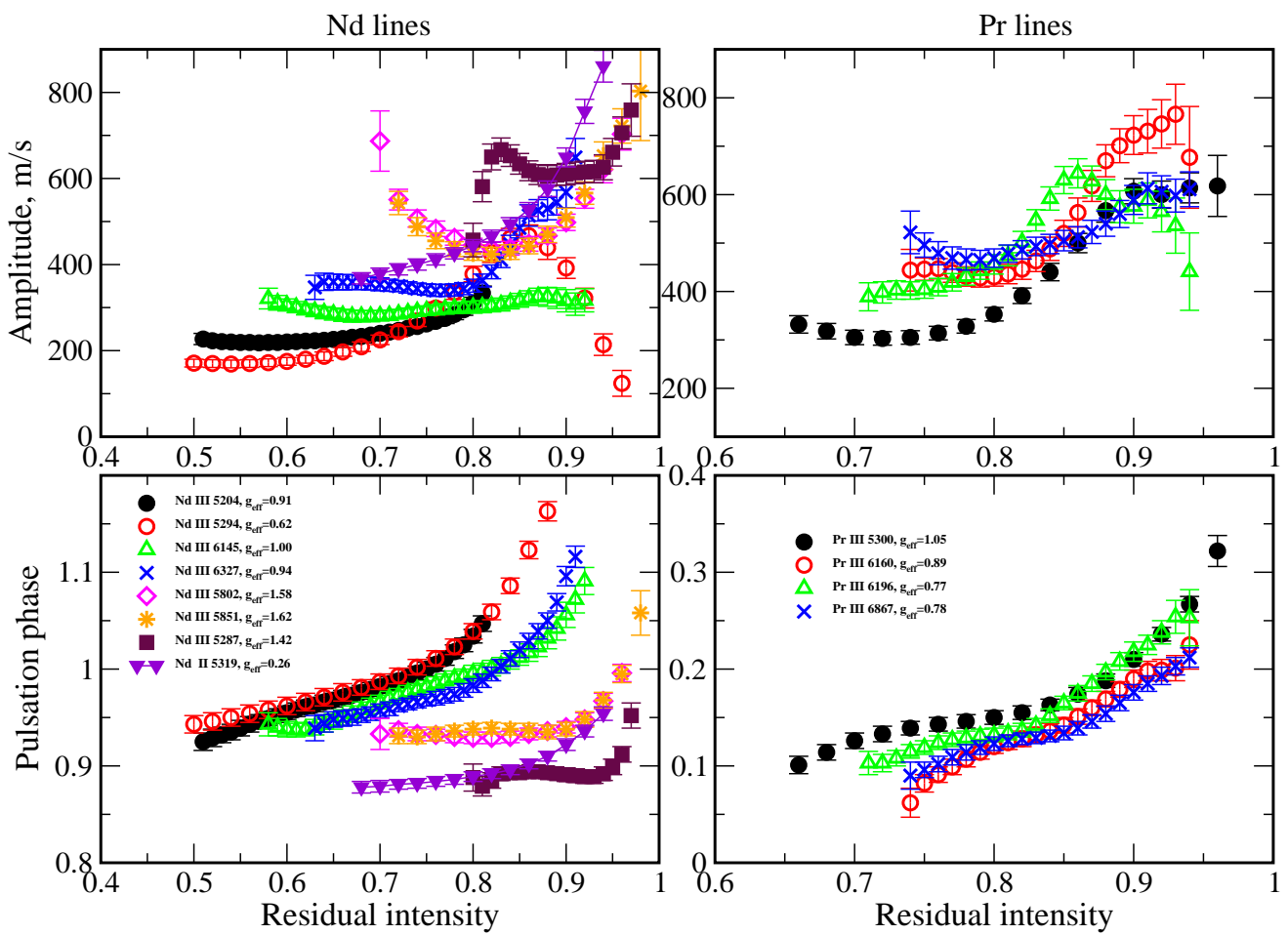}
\caption{Bisector measurements as a function of the Nd (left panels)
and Pr (right panels) lines' residual intensity in HD~134214. Top
panels represent RV amplitudes, and bottom panels show pulsation
phases calculated with the main period from Table~\ref{tbl2}.}
\label{HD134214_Nd}
\end{figure*}

More care should be taken in the choice of metal spectral lines for
the bisector measurements than for the centre-of-gravity RV
analysis. Blends with non-pulsating lines change the run of bisector
pulsation amplitude across the line profiles, while blends with
pulsating lines (weak REE lines in far wings, for example) may
change both amplitudes and phases. These artifacts often lead to the
wrong conclusion about roAp pulsational behaviour, especially when
these conclusions are based on analysis of only one line. For
instance, the well-known Nd\iii\, 6145.068~\AA\ line, analysed in
many pulsation studies, is blended with the Si\i\, 6145.016~\AA\
line and with the Ce\ii\,~6144.853 \AA. The latter feature makes a
non-negligible contribution to the spectrum of Przybylski's star.
The strongest Nd\iii\, 5294.11~\AA\ line is blended with the
Fe\i\,~5293.96~\AA\ line in the blue wing and with the
high-excitation Mn\ii\, 5294.32~\AA\ line in the red wing. Both
lines are normally weak, but they are strengthened in the stratified
atmospheres of roAp stars with the effective temperatures higher
than 7000--7200 K. This blending results in a drop of the bisector
velocity amplitudes starting from some intensity points in the line
profile. Zeeman splitting may also change velocity amplitude
distribution across the line profile. Figure~\ref{HD134214_Nd}
illustrates the influence of blends and Zeeman effect on the
bisector amplitudes and phases of Nd\ii-\iii\ and Pr\iii\ lines in
HD~134214. Pairs of Nd\iii\ lines with similar intensities and
Zeeman structure have identical bisector phases independent of any
blending. Thus, phases may be considered as a more reliable
indicator of the pulsational characteristics. One may see that, up
to the residual intensity 0.85, the Nd\iii\,5294~\AA\ line shows the
same bisector velocity amplitudes as another strong Nd\iii\ line at
5204~\AA. However, after this intensity point, the two amplitude
curves start to deviate: in the Nd\iii\, 5294~\AA\ line, the
amplitude drops towards the line wings, probably due to blends. Also
blends decrease bisector amplitudes in the Nd\iii\,6145~\AA\
compared to the Nd\iii\,6327~\AA\ line, which has the same intensity
and Zeeman structure. Three out of the four Pr\iii\, lines shown in
Fig.~\ref{HD134214_Nd} are practically free of blends in most of the
programme stars, and additionally, these lines have identical Zeeman
splitting, which results in similar bisector pulsational behaviour.

\begin{figure*}[!th]
\centering
\firrps{82mm}{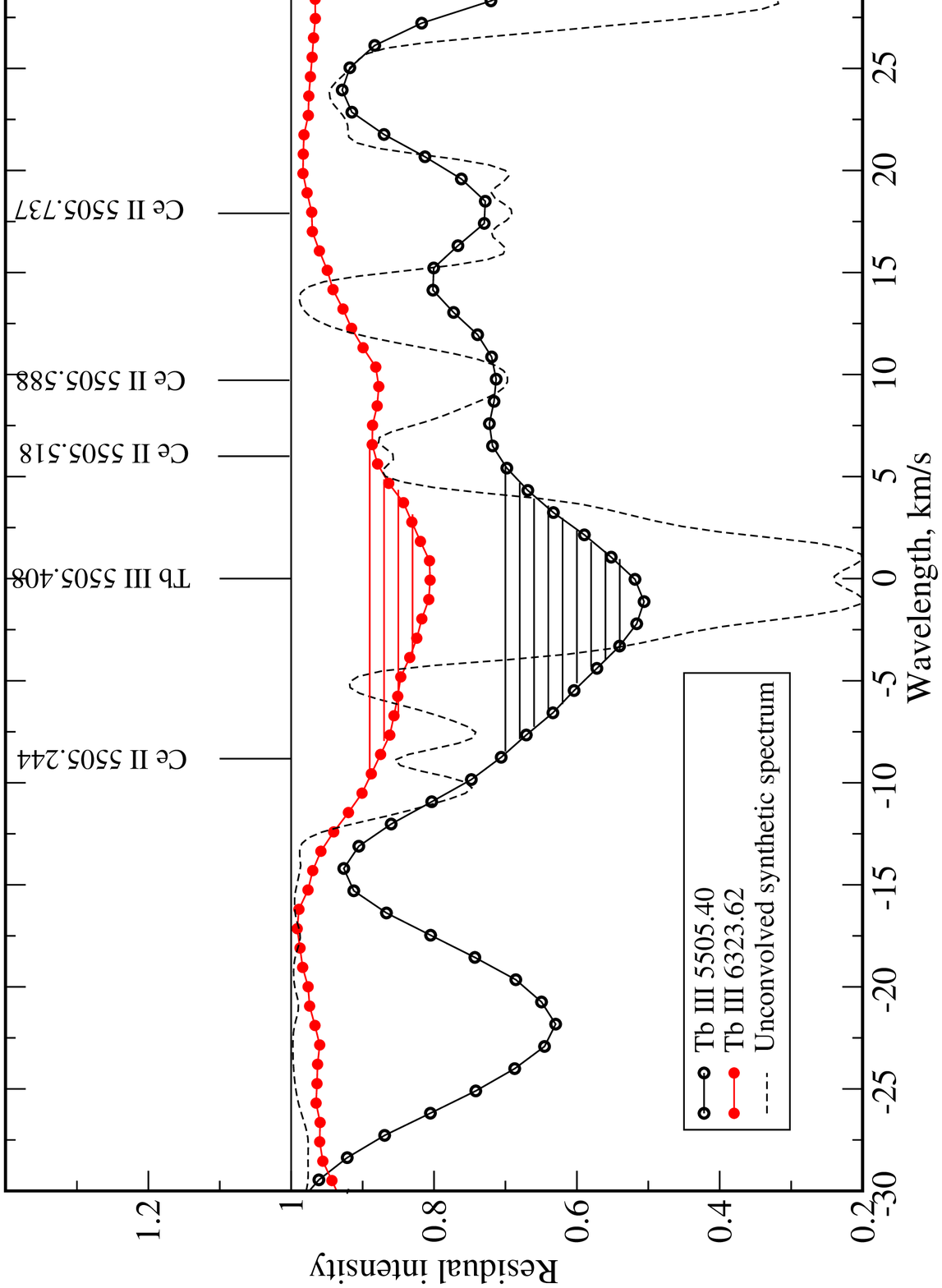}\hspace{0.5cm}\firrps{82mm}{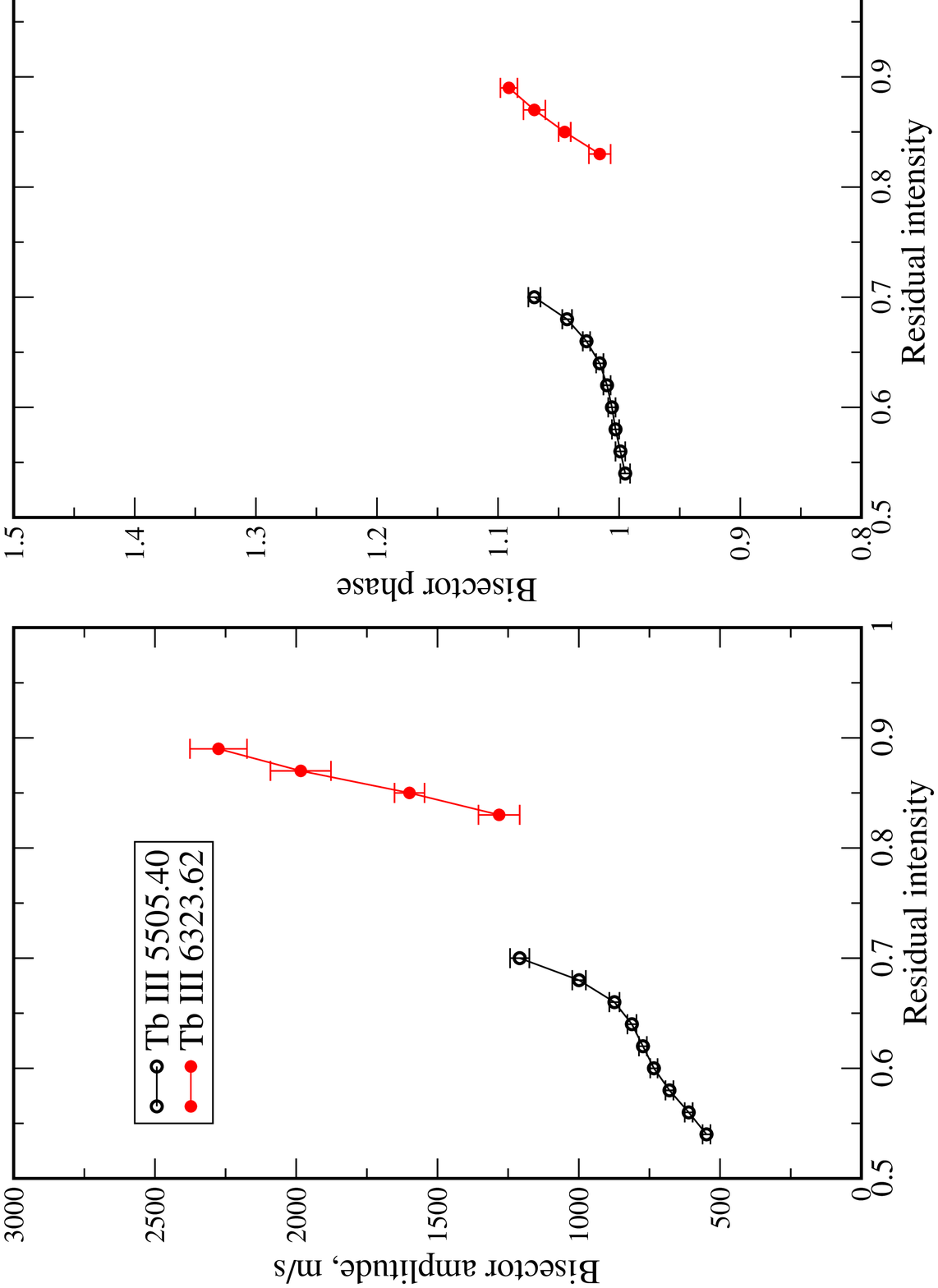}
\caption{Tb\iii\, lines in HD~101065. Left panel represents the
observed and calculated line profiles, and the two next panels show
the RV amplitudes and pulsation phases determined for the intensity
levels indicated in the corresponding line profiles by the
horizontal lines.} \label{Tb3}
\end{figure*}

As shown in Section\,\ref{ph-a}, Tb\iii\ lines exhibit a
particularly interesting pulsation behaviour, so their blending should
be investigated in detail. The Tb\iii\ lines are not as strong and
numerous as Pr\iii\ and Nd\iii. From 10 lines only the strongest
one, Tb\iii\ 5505 and three others, 5847, 6323, and 6687~\AA, may be
used for bisector measurements. Most often we measured the Tb\iii\
5505.408~\AA\ line. This line is blended by the Ce\ii~5505.204~\AA\
in the blue wing and by Ce\ii~5505.588~\AA\ in the red wing. The
Tb\iii~$\lambda$~6323.619~\AA\ line is free of blends but may be
partially blended with the weak atmospheric O$_2$ line at $\lambda$
6323.75~\AA. Figure~\ref{Tb3} shows a spectral region around the
Tb\iii\ 5505.408~\AA\ line in HD~101065, where the blending problem
is the most severe. For comparison we show the region with the
Tb\iii\ 6323.619~\AA\ line; therefore, wavelength scale is given in
\kms\ relative to the centre of each line. We also show synthetic
spectrum calculated with the highest possible Ce abundance and
magnetic field strength appropriate for HD\,101065 (see
Table~\ref{tbl2}). Although the full intensity of the calculted
feature is equal to the intensity of the observed one, for
demonstration of the blending effects we did not convolve synthetic
spectrum with the instrumental and rotational profiles. Note that
the red wing of the Tb\iii\ 6323.619~\AA\ line is blended with the
atmospheric O$_2$ line 6323.75, therefore no bisector measurements
were carried out above intensity level 0.9. The blue wing asymmetry
in both Tb\iii\, lines is rather an effect of the hyperfine
splitting, than the real influence of blends, because the same
asymmetry is observed in other stars where the Ce\ii\, contribution
is negligible.

Taking into account the
similarity of the RV amplitudes and, in particular, phases in the bisector measurements of
both lines we conclude that even in HD~101065 the blending does not affect seriously the
measured pulsation characteristics.

\section{Phase-amplitude diagrams}
\label{ph-a}

\begin{figure*}[!th]
\centering
\firrps{82mm}{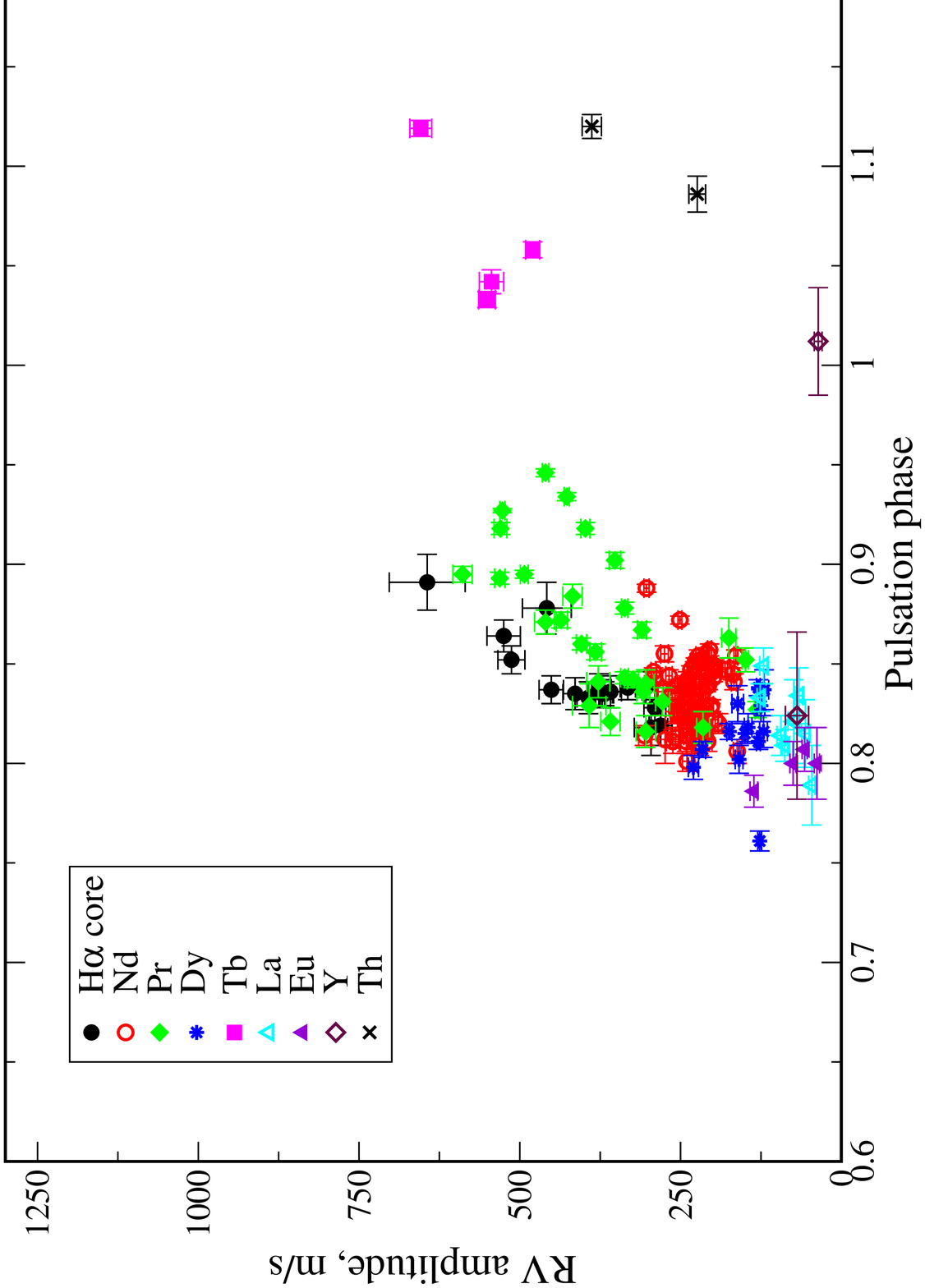}\hspace{0.5cm}\firrps{82mm}{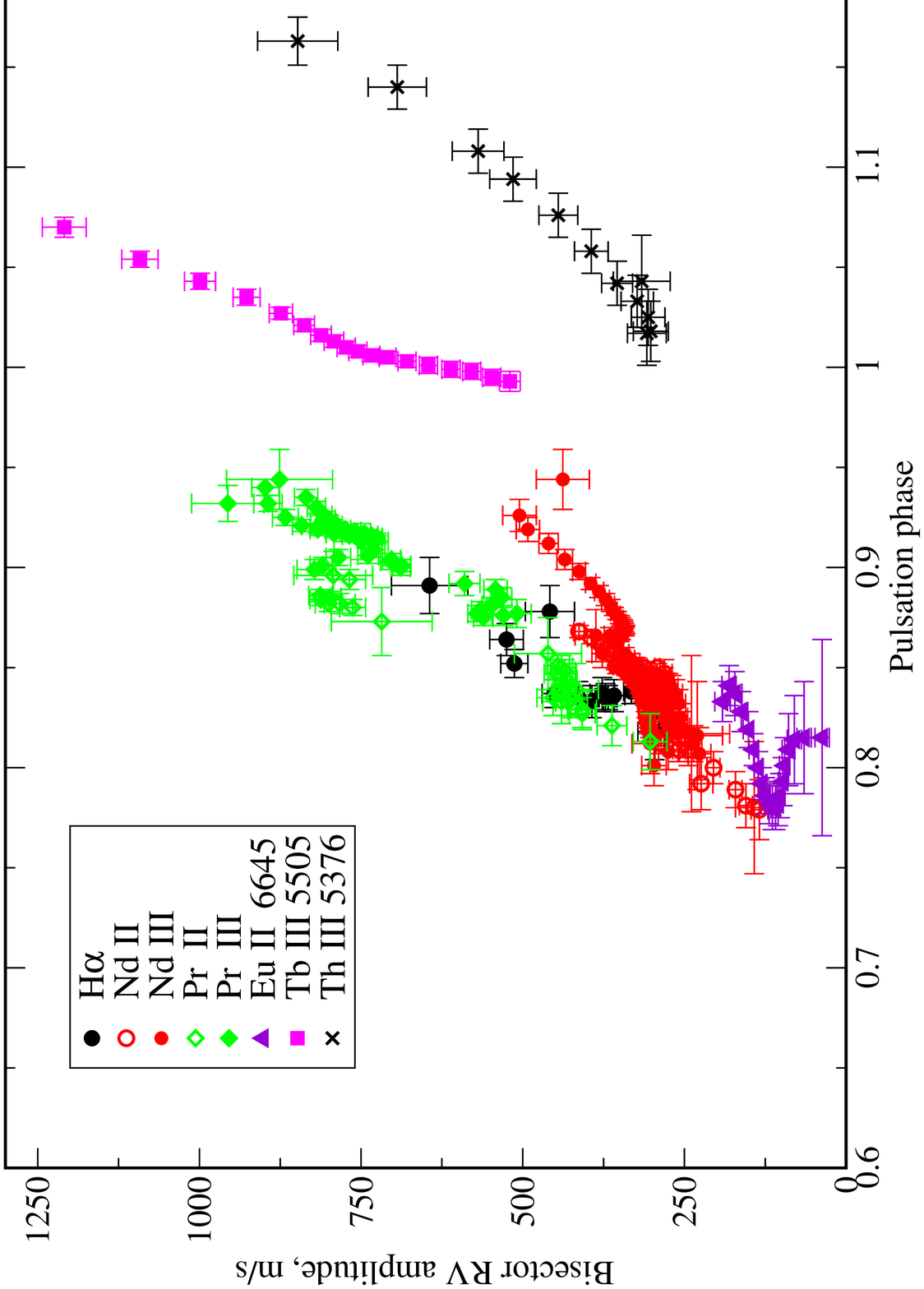}
\caption{Amplitude-phase diagrams for the pulsation centre-of-gravity measurements (left panel) and bisector measurements (right
panel) in HD~101065.}
\label{HD101065}
\end{figure*}

Empirical understanding and detailed theoretical interpretation of
the pulsation phenomenon in roAp stars requires construction of the
vertical cross-section of pulsation modes. Chemical stratification
enables this difficult task by separating formation layers of the
spectral lines of different chemical elements far apart, thus
allowing us to resolve the vertical dimension of pulsating Ap-star
atmospheres, as can be done for no other type of pulsating stars.
However, the key information needed for such vertical pulsation
tomography -- formation depth of REE spectral lines -- is difficult
to obtain. Early studies (e.g., Kanaan \& Hatzes \cite{KH98}) have
suggested that the line intensity may be used as a proxy of the
relative formation heights. However, it is now understood that the
chemical stratification effects  are dominant in the atmospheres of
cool Ap stars and, therefore, formation region of weak lines of one
element is not necessarily located deeper than the layer from where
the strong lines of another element originate. This is why a
comparison of the intensities of different pulsating lines is only
meaningful for the absorption features of the same element/ion.

The study by Ryabchikova et al. (\cite{RPK02}) presented the first
comprehensive analysis of the vertical stratification of light,
iron-peak, and rare-earth elements in a roAp star, demonstrating
that only by taking height-dependent abundance profiles into account
can one calculate correct formation depths of pulsating spectral
lines and meaningfully interpret the results of time-resolved
spectroscopy. In a later paper, Mkrtichian et al.
(\cite{MHK03}) study the vertical profiles for $p$-mode oscillations
in 33~Lib (HD\,137949) using pulsation centre-of-gravity
measurements for individual lines. To establish the vertical
atmospheric coordinate, the authors used the $W_{\lambda} - \log
\tau$ transformation scale in 33~Lib's atmosphere by assuming
homogeneous elemental distribution. This simplified approach was
sufficient for detecting the node in the Nd\ii, Nd\iii\ line-forming
region -- a discovery later confirmed by Kurtz et al.
(\cite{KEM05b}) and by the present paper. However, as Mkrtichian et
al. (\cite{MHK03}) themselves acknowledge, their method is too
simplified because it ignores the vertical and horizontal
inhomogeneous distribution of chemical abundances. Indeed, their
technique cannot be used for simultaneous interpretation of the
pulsational variability of different elements since Fe, as well as
Ca, Cr and all REE elements, are strongly stratified in the
atmosphere of 33~Lib and other roAp stars (see Ryabchikova et al.
\cite{RNW04}), and stratification is not the same for different
elements. Mkrtichian et al. (\cite{MHK03}) suggest that in future
studies some of these difficulties can be alleviated by the
``multi-frequency tomographic approach'' that will use the phase and
amplitude profiles of different oscillations modes to establish a
link to the atmospheric geometric depth scale.

In general, the detailed pulsation tomography analysis of roAp stars
should be based on sophisticated atmospheric modelling  including
chemical stratification, NLTE, magnetic field, and eventually,
pulsation effects on the shape and intensity of spectral lines with
substantial RV variability. Such modelling is very demanding in
terms of the quality of observations, required input data, and
computer resources. This is why only two roAp stars, $\gamma$~Equ
and HD\,24712, have been studied with this method up to now
(Mashonkina et al. \cite{MRR05}; Ryabchikova et al. \cite{RMR06}).
Here we suggest a different approach to the pulsation tomography
problem. In the framework of the outward propagating magnetoacoustic
wave, one expects a continuous amplitude versus phase relation for
pulsation modes. Lines showing later RV maximum should originate
higher in the atmosphere.  Thus, new insight into the roAp pulsation
modes structure can be obtained by inferring and interpreting  the
trend in pulsation velocity amplitude as a function of pulsation
phase. Such phase-amplitude diagrams offer a possibility of tracing
the vertical variation in the mode structure without tedious
assignment of the physical depth to each pulsation measurement.

We note that this paper is not the first one to use the
phase-amplitude diagrams for roAp stars. Baldry et al. (\cite{BB98})
and Baldry \& Bedding (\cite{BB00}) produced similar diagrams in the
pulsational studies of $\alpha$~Cir and HR~3831. However, their
diagrams described pulsational properties not for individual lines
with a proper identification but for small spectral regions often
containing several spectral lines, sometimes without correct
identification. As a result, the most important information on wave
propagation was lost (see Sect. ~\ref{alcir}).

For each programme star we have produced phase-amplitude diagrams
using a set of lines of representative chemical species, including
Y\ii, Eu\ii, the H$\alpha$ core, La\ii, Dy\ii, Dy\iii, Nd\ii,
Nd\iii, Pr\ii, Pr\iii, Tb\iii, and Th\iii. We considered velocity
amplitudes and phases derived with both the centre-of-gravity RV
measurements and the bisector analysis. The line blending varies
significantly from star to star due to different effective
temperatures and chemical anomalies, therefore we cannot use
identical set of spectral lines for all stars. For each star we
tried to employ unblended or minimally blended lines, in particular,
for the bisector measurements. In addition, the line broadening
expressed in the terms of macroturbulent velocity was estimated for
a few representative lines of each chemical species. This
information is useful for assessing the isotropic velocity component
at the formation heights of REE lines.

Below we present the results for individual stars.

\subsection{HD~101065 (Przybylski's star)}



The atmosphere of HD~101065 is known to be very rich in REE and
underabundant in most other elements, including the Fe-peak species.
Due to the low effective temperature, slow rotation, and abundance
anomalies, most lines in the spectrum of HD~101065 are strong and
sharp. As a result, one can achieve impressive accuracy of pulsation
measurements, typically $\sim$5--8~\ms, for moderately strong lines.
Although it is not easy to find unblended Fe lines in the forest of
the REE lines, we managed to measure a few of them and to detect no
pulsational variability above 30~\ms. Pulsation amplitudes at the
level of 6 to 17~\ms\ were detected in the cores of very strong
Ba\ii\, $\lambda\lambda$~6142, 6496~\AA\ lines, which gives us an
idea about pulsation amplitudes close to the photospheric layers.

The centre-of-gravity and bisector phase-amplitude diagrams for
HD\,101065 are shown in Fig.~\ref{HD101065}. The centre-of-gravity
measurements give us a general idea of the pulsation wave
propagation in roAp atmosphere. Pulsations are characterised by the
amplitude $\sim$40~\ms\ at the levels of the Y\ii, La \ii, and
strong Eu\ii\ line formation. Then they pass, with the gradually
increasing amplitude, the layers where the Dy\ii, Dy\iii, Nd\ii,
Pr\ii, Nd\iii, the H$\alpha$ core and Pr\iii\ lines are formed. The
phase does not change by more than 0.15 of the pulsation period
(less than 1 radian) from Eu\ii\, to Pr\iii\, lines. After that, a
rapid change in pulsation phase occurs in Tb\iii\, and, next, in
Th\iii\, lines. We measured two unblended Th\iii\,lines at
$\lambda$~5376.13 and 6599.48~\AA.

HD~101065 is the only star in our sample that shows similar
pulsation signatures (bisector amplitude versus pulsation phase) for
the \ha\ core, Nd\ii, Nd\iii, Pr\ii\, and weak Pr\iii\, lines. For
all these features we find an increase in both pulsation amplitude
and phase from the line wings to the line core. Kurtz et al.
(\cite{KEM05a}) have obtained similar amplitudes and phases for the
H$\alpha$ core and, based on the bisector measurements of the single
Nd\iii\, 6145~\AA\ line, they argue that the formation layers of
Nd\iii\, lines start above the layers of formation of the deepest
part of the H$\alpha$ core. Our Fig.~\ref{HD101065} provides strong
evidence that most of REE lines, including Pr\ii\, and weaker
Pr\iii\, lines, are formed in the same layers as the H$\alpha$ core.

In addition we analysed the broadening of pulsating lines. All
low-amplitude Ce\ii\, lines do not exhibit any extra broadening
above the adopted value of the projected rotation velocity. All
Eu\ii\ lines and Dy and Nd lines in both ionisation states show an
extra broadening corresponding to \vmacro\, between 2.5 and 4~\kms.
Macroturbulent velocity grows from 5~\kms\, in weak Pr\iii\,
5765~\AA\ line to 7--8~\kms\, in the strong Pr\iii\, lines
$\lambda\lambda$~5300, 6707~\AA. We need \vmacro=10~\kms\ to fit the
observed profiles of the unblended Tb\iii\ 6832~\AA\ and Th\iii\
5376~\AA\ lines. It is worth noting that this rapid increase in the
line broadening accompanies the change in the line profile
variability pattern from the normal symmetric shape to the
blue-to-red running waves, which become noticeable for the strongest
Pr\iii\ lines and, in particular, for Tb\iii\, lines (Kochukhov et
al. \cite{KR07}). An extra broadening of the Pr\iii\, lines cannot
be attributed to hyperfine structure.

Considering the amplitude-phase diagrams, we see a change in the
pulsation behaviour for Tb\iii\, lines and, even stronger, for the
Th\iii\ 5376.13~\AA\ line. In contrast to the variability of the
H$\alpha$ core and Nd lines, we observe a rapid increase in the
amplitude and in the phase from the line core to line wings in the
lines of Tb\iii\, and Th\iii\, (see Fig.~\ref{HD101065_prof} - Online only). This
phenomenon cannot be explained by the pulsation wave propagating
outwards.

In general, over the large fraction of the Przybylski's star
atmosphere, up to the layers of the Nd\iii\, and weak Pr\iii\, line
formation, pulsations are represented well by non-radial
non-adiabatic {\it p}-modes (Saio \cite{S05}) with an amplitude
increasing outward.  Above these layers, pulsation characteristics
change in a way that is not forseen by the theoretical models.

\subsection{HD~122970}

\begin{figure}[!t]
\centering
\firrps{82mm}{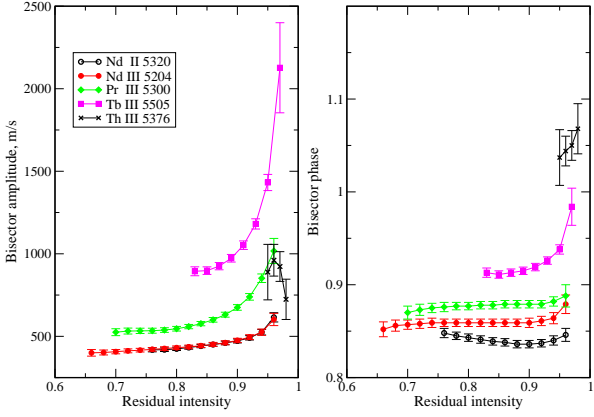}
\caption{Bisector measurements as a function of lines residual
intensity in HD~122970. The left panel represents RV amplitudes,
right panel shows pulsation phases calculated with the main period
from Table~\ref{tbl2}.} \label{HD122970_int}
\end{figure}

Pulsational observations of this star were obtained at the
rotational phase 0.75, between the maximum and minimum of the
magnetic field (see Fig.~3 in Ryabchikova et al. \cite{RWA05}). The
centre-of-gravity and bisector phase-amplitude diagrams for
HD\,122970 are presented in Fig.~\ref{HD122970} (Online only). For a few chosen
lines we also show the variability of the pulsational
characteristics along the line profiles in Fig.~\ref{HD122970_int}.

The overall pulsational behaviour in the atmosphere of this star is
similar to HD~101065: low-amplitude pulsations are seen in Y\ii\,
and Eu\ii\, lines, the amplitude increases slightly in the layers
where the H$\alpha$ core forms, and then it increases further in the
Dy and Nd line formation layers. A rapid growth of the amplitude
occurs in the region of the Pr line formation. Up to these layers,
pulsations have a standing-wave character, with almost constant
phases for all lines of Y\ii, Eu\ii, La\ii, Dy, Nd, and Pr. The
phase shifts become noticeable for Tb\iii\, and Th\iii\, lines, as
in HD~101065, and, again, in these lines pulsation amplitude and
phase increase from the line core to line wing, contrary to
predicted effect of the outward-running wave. While in HD~101065 the
regions of the H$\alpha$ core, Nd, and Pr line formation are the
same or, at least, partially overlap, in HD122970 the
phase-amplitude diagrams show a significantly more stratified
atmospheric structure.

Ryabchikova et al. (\cite{RSH00}) noted that the projected
rotational velocities derived from the REE lines are lower than
those obtained from Fe-peak lines. More precise present observations
confirm this result. We determined \vs\,$\approx$\,3.5~\kms from the
REE line fitting, whereas the Fe-peak elements show broadening
corresponding to \vs\,=\,4.5~\kms. A 3.877~d rotation period
obtained by Ryabchikova et al. (\cite{RWA05}) for HD\,122970,
together with the low \vs, suggests a nearly pole-on geometry.
Therefore, the difference in the apparent rotational velocities of
the Fe-peak and REE lines may indicate a concentration of the REEs
near the visible stellar rotation pole. No extra broadening is
required to fit the Nd\ii, Nd\iii, Pr\ii, and weak Pr\iii\ lines,
while \vmacro\,=\,2--3~\kms\, and 6 \kms\, are needed to fit the
strong Pr\iii\ lines and the Th\iii\ 5376~\AA\ line, respectively.
The Y\ii\ lines, which show the smallest detectable RV amplitudes,
have the same pulsation phases and extra broadening as strong
Pr\iii\ lines.

The lines of the elements with Z\,$\le$\,38 do not show any pulsation amplitude above 7--20~\ms. Two measured Zr\ii\, lines have amplitudes
of $\approx$\,200~\ms.

\subsection{HD~24712}




According to the ephemeris given by Ryabchikova~et~al.
(\cite{RWA05}), our pulsational observations were obtained at the
rotational phase 0.944, i.e. close to the magnetic maximum. Detailed
observational analysis of the spectroscopic pulsation signatures in
HD~24712 was recently presented by Ryabchikova et al.
(\cite{RSW06}). HD~24712 is the only roAp star for which NLTE line
formation calculations of \ha, Pr, and Nd were carried out
(Mashonkina et al. \cite{MRR05}; Ryabchikova et al. \cite{RMR06}).
The observed intensities of the Nd and Pr lines in the first and the
second ionisation states are explained by the stratified atmosphere
with a step-like enhancement of these elements  just above the NLTE
formation depth of the \ha\ core ($\log\tau_{5000}$\,=\,$-4$). Up to
this atmospheric level, the observed distribution of pulsation
phases agrees rather well with the predictions of the non-adiabatic
pulsation models (see Fig.~1 in Kochukhov \cite{K06} and Fig.~3 in
Sachkov et al. \cite{SR06}), which is supported by the
amplitude-phase diagrams for HD~24712 shown in Fig.~\ref{HD24712} (Online only).
In contrast to the standing-wave dominated behaviour of HD~101065
and HD~122970, we see a running wave in HD~24712 from the low
atmosphere, where the first pulsation signal is detected. Due to the
availability of the simultaneous spectroscopic and photometric
monitoring, the sequence of the pulsation phase changes is
established with high accuracy in this star (Ryabchikova et al.
\cite{RSW06}). Pulsation follows the same order as in HD~101065 and
in HD~122970: Eu, La, \ha\ core, Nd, Pr, Tb lines.

The phase distribution with the optical depth, as well as our
amplitude-phase diagrams, shows the phase shifts between \ha\, and
Nd lines and between Pr and Nd lines. The NLTE calculations do not
predict a significant difference in the location of the Pr and Nd
line formation regions, therefore the observed phase shift is
probably caused not by the vertical distribution of chemical
elements, but by the horizontal abundance inhomogeneities. According
to the pulsation phases, Nd lines have to be formed at the same
layers as the deeper parts of the \ha\ core. At the same time, we
caution that the existing NLTE calculations of \ha, Pr, and Nd lines
are very preliminary, and they do not include potentially important
effects of stratified abundance distribution on the atmospheric
structure and neglect an influence of the magnetic desaturation on
the line formation.

The bisector velocity amplitudes of HD\,24712 are either constant
along the line profile, or they slightly increase from the line
wings to the line core, except for the Tb\iii\ and, possibly, Y\ii\
lines, in which bisector amplitude increases from the line core to
the line wings (Fig.~\ref{HD24712_int} - Online only). Pulsation phases are nearly
constant along the profiles of most lines, but show a tendency to
grow from the line centre to the line wings for the strongest
Pr\iii\ lines and for Tb\iii\ lines. Both Th\iii\ lines show
pulsational amplitudes close to the detection limit. A very
interesting result is obtained for Y\ii\, lines. Yttrium is the only
element with Z\,$\le$\,40 whose lines have a measurable pulsation
amplitude in HD\,24712 (see Ryabchikova et al. \cite{RSW06}). These
lines are strong enough to provide precise centre-of-gravity and
bisector measurements, therefore pulsation phases can be determined
with high accuracy. While RV amplitudes do not exceed 50--100~\ms\
for the strongest Y\ii~5087~\AA\ line, the bisector phase
distribution across the line profile coincides with the phase
distribution across the profile of the strongest Pr\iii\ 5300~\AA\
line, which has an RV amplitude above 300~\ms.  The estimated depth
of formation of Y\ii\ lines in chemically homogeneous atmosphere
lies around $\log\tau_{5000}=-1.5$ to $-2.5$, where the core of \ha\
line starts forming. Note that Doppler imaging of HD~24712 shows a
similarity between yttrium and REE surface distributions
(L\"uftinger et al. \cite{MDI06}). The observed phase difference
between Y\ii\ and the \ha\ and the coincidence of the phases in
Y\ii\ and strong Pr\iii\ lines, on the one hand, and the large
difference in amplitudes, on the other, is difficult to explain. We
will see that the same pulsational behaviour of Y\ii\ lines is
observed for some other roAp stars.

Although the projected rotational velocity of HD~24712 is rather
high, spectral synthesis clearly shows that we need an extra
broadening, equivalent to \vmacro\,=\,6~\kms, to fit strong Pr\iii\
lines, as well as Tb\iii\ and Th\iii\ lines, whereas no extra
broadening is required for Nd lines. Interestingly, one needs
\vmacro\,=\,4~\kms\ to reproduce the profile of the Y\ii~5087~\AA\
line.

\subsection{HD~134214}
\label{hd134214}

This star has one of the shortest pulsation periods, 5.7 min, which
is close to the acoustic cut-off frequency calculated for HD~134214
by Audard et al. (\cite{AKM98}). The centre-of-gravity and bisector
amplitude-phase diagrams are shown for HD\,134214 in
Fig.~\ref{HD134214} (Online only). The amplitude-phase diagrams for this star are
similar to those for HD~24712. In both stars we observe the running
wave from the regions of Eu\ii\, line formation to Tb\iii, Th\iii\,
line-forming regions. In HD~134214, all lines pulsating with high
amplitude show an increase both in amplitude and in phase from the
line core to the line wings (see Fig.~\ref{HD134214_Nd}).

The low rotational velocity allows us to partially resolve
components of the Zeeman split lines, which makes the line profile
analysis easier and, in particular, allows us to estimate an extra
broadening accurately. An extra broadening of $\sim$1~\kms\, is
needed to fit the low-amplitude pulsating lines of Eu\ii. Then the
inferred \vmacro\, increases gradually: 4~\kms\, (Nd\ii, Dy\iii),
5--6~\kms\, (Nd\iii), 8--10~\kms (Pr\iii), 10~\kms\, (Tb\iii), and
more than 10~\kms\, (Th\iii). Just as in HD~24712, Y\ii\, lines have
small detectable RV amplitudes, but the same bisector phases as
strong Pr\iii\, lines do. We measured \vmacro\,$\sim$6~\kms\ for the
Y\ii\ 5087~\AA\ line.

In both stars the bisector amplitude-phase diagrams for Nd\ii\, and
weaker Nd\iii\, lines are overlapping, which supports the model of
the Nd stratification in roAp atmospheres proposed by Mashonkina et
al. (\cite{MRR05}). According to this model the strongest lines of
Nd\ii\, are formed at the same layers as weak Nd\iii\, lines,
therefore their pulsational characteristics should be similar.

No definite pulsation signatures are detected in
the lines with Z\,$\le$\,38 and also in  Zr\ii\, and Ba\ii\, lines. The upper RV amplitude limit is 10~\ms.

\subsection{HD~128898 ($\alpha$~Cir)}
\label{alcir}

\begin{figure}[!t]
\centering
\firrps{82mm}{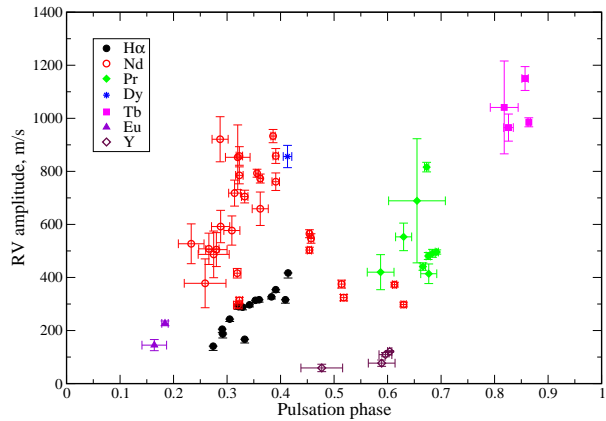}
\caption{Amplitude-phase diagrams for the pulsation centre-of-gravity measurements  in $\alpha$~Cir.}
\label{alcir_diagr}
\end{figure}

The main photometric period of $\alpha$~Cir, one of the short-period
roAp stars, is 6.83 min, which is close to the acoustic cut-off
frequency. The amplitude-phase diagram for HD\,128898
(Fig.~\ref{alcir_diagr}) appears to be similar to those constructed
for HD~24712 and HD~134214 and  indicates a running wave. Based on
an amplitude-phase diagram made for short spectral regions
containing groups of unresolved lines, Baldry et al. (\cite{BB98})
suggested a high-overtone standing wave with a velocity node in the
atmosphere of $\alpha$~Cir. This contradicts our results obtained
from analysis of pulsations in isolated lines. Again, similar to
HD~24712 and HD~134214, Y\ii\ lines have the smallest RV amplitudes,
but the same phases as the strongest Nd\iii\ and Pr\iii\ lines.

The bisector analysis shows that bisector amplitude rapidly
decreases from the line centre to the line wings with small changes
in phase. This behaviour may be explained by a combination of
non-radial pulsations and surface chemical inhomogeneity, which
clearly manifests itself in the rotational modulation of the line
profiles in $\alpha$~Cir (Kochukhov \& Ryabchikova \cite{KR01b}).
The average spectrum used in the present study corresponds to the
rotation phase 0.44 (minimum of the photometric pulsational
amplitude) if we apply the ephemeris from Kurtz~et~al.
(\cite{acir}). All REE lines in our $\alpha$~Cir spectrum exhibit
two-component profiles, which represents another proof 
of chemical spots.

\subsection{HD~12932}
\label{HD12932}

Kurtz et al. (\cite{KEM06}) suggest that HD~12932 is an example of
the star with a standing wave behaviour in the layers of the \ha\
core formation based on the constancy of the bisector pulsation
phases. Our measurements are presented in Fig.~\ref{hd12932} (Online only). While
for Nd\ii\ and the \ha\ core the bisector phases are almost constant
for a given line, the phases differ from line to line. Phase shifts
are small, but reliable, taking the high accuracy of the phase
determination into account (Fig.~\ref{HD12932_int} - Online only). The bisector
phase measurements show that, if we attribute a given phase to the
particular atmospheric height, many Nd lines should originate in
extremely thin separate layers, which is impossible to explain.



\begin{figure}[!t]
\centering
\firrps{82mm}{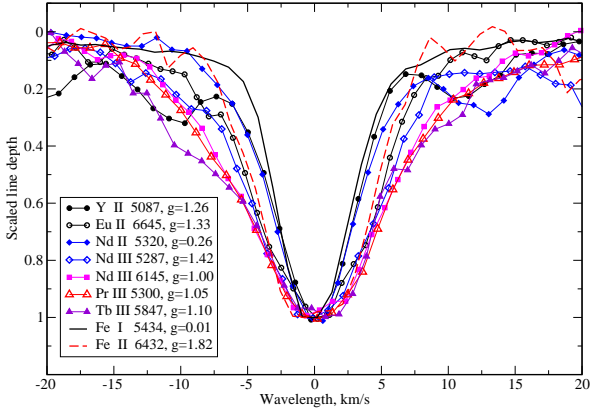}
\caption{A comparison of the line profiles of different elements in HD~12932. Line profiles are scaled to the same central depth.
}
\label{hd12932_prof}
\end{figure}

Due to the slow rotation and a rather weak magnetic field, HD~12932
is useful for demonstrating the effect of differential extra
broadening of the pulsating lines.  Figure~\ref{hd12932_prof}
presents a comparison between the line profiles of different
elements/ions, scaled to the same central line depth. The wavelength
scale is given in \kms, with line centre at zero velocity. The two
iron lines with zero and large Zeeman broadening are shown to assess
the expected magnetic field broadening effect. These lines,
Fe\i~5434~\AA\ and Fe\ii~6432~\AA, do not require any additional
broadening other than Zeeman, instrumental, and rotational ones to
fit the observed line profiles. The Eu\ii\ line broadening is
explained by the combined Zeeman and hyperfine splitting. Our
spectral synthesis shows a growth in the \vmacro\ from Nd\ii\, lines
(3--5 \kms), to Nd\iii\ lines (6--9 \kms), Pr\iii\ lines (7--10
\kms), and finally, to Tb\iii\ lines (10--11 \kms). The stronger
line of the same ion, the larger \vmacro\ is required to fit the
line profile. The Th\iii\ lines in HD\,12932 are too weak for
reliable measurements. The Y\ii~5087~\AA\ line exhibits an extra
broadening similar to Nd\ii\ lines, although no definite pulsation
signal was detected in all measured Y\ii\ lines, as well as in the
lines of Na, Mg, Si, Ca, Sc, Cr, Fe, Co, Ni, and Sr. No variability
is detected in the Fe\i\ 5434~\AA\ line (RV amplitude 7$\pm$15~\ms).
Thus, we do not confirm results by Kurtz et al. (\cite{KEM05a}), who
claim a 3$\sigma$ detection of pulsations in this line. We looked at
47 other lines of Fe\i\ and Fe\ii, and we detected a pulsation
signal exceeding the 3$\sigma$ significance level only in seven
features. For most of these lines weak variability is explained by
blending with weak REE lines. Three out of six measured Ti\ii\ lines
show a pulsation signal with the RV amplitudes 80--100~\ms, while
two T\i\ lines did not reveal any variability. All three Ba\ii\
lines show RV amplitudes of $\approx$200~\ms.

\subsection{HD~201601 ($\gamma$~Equ)}
\label{gequ}

The amplitude-phase diagrams for the centre-of-gravity measurements
in $\gamma$~Equ are shown in Fig.~\ref{gequ-rv} (Online only; left panel). Rather
low spectral resolution of the $\gamma$~Equ observations used for
this type of pulsation measurements does not allow us to make a
precise bisector study. For the bisector analysis we used the
time-series observations of $\gamma$~Equ carried out with the Gecko
spectrograph at CFHT (see Kochukhov et al. \cite{KR07}). The
bisector amplitude-phase diagrams are displayed in
Fig.~\ref{gequ-rv} (right panel). If in the previous star, HD~12932,
the RV amplitude remained nearly constant above a certain atmospheric
height, in $\gamma$~Equ we observed a continuous decrease of RV
amplitude outward. The bisector amplitude-phase diagrams are fairly
similar in both stars, showing the same constancy in the phase
measured across Nd spectral lines and a phase shift between lines
(Fig.~\ref{HD201601_int} - Online only). This phase shift is larger in
$\gamma$~Equ. Again, Y\ii\ lines have very small RV amplitudes, but
their pulsation phases are the same as in the strongest Nd\iii\ and
Pr\iii\ lines. Extra broadening of the REE lines in $\gamma$~Equ was
discussed by Kochukhov \& Ryabchikova (\cite{KR01a}), who showed
that \vmacro\,=\,10~\kms\ is needed for fitting strong Nd\iii\ and
Pr\iii\ lines.



\subsection{HD~19918}
\label{HD19918}

Pulsational characteristics of HD~19918 (Fig.~\ref{hd19918-rv}) are
similar to both HD~12932 and, in particular, to $\gamma$~Equ. An
even more rapid decrease in the RV amplitude is observed in the
atmosphere above a certain layer. The three stars show the largest
RV amplitudes and the largest variation in the bisector velocity
amplitudes from the line centre to the line wings
(Fig.~\ref{HD19918_int} - Online only). In all these stars we find first an
increase in the RV amplitudes with approximately constant phases
(standing wave), and after that the pulsation wave transforms into a
running one.

As in HD~12932, differential extra broadening is required to fit
line profiles: \vmacro\,=\,0~\kms\ (Fe), \vmacro\,=\,2--3~\kms\
(Y\ii),  \vmacro\,=\,4~\kms\ (Eu\ii), \vmacro\,=\,7--8~\kms\
(Nd\ii), \vmacro\,=\,8--10~\kms\ (Nd\iii\ and Pr\iii), and
\vmacro\,=\,11~\kms\ (Tb\iii). Note that a smaller broadening is
necessary to fit the line core than the line wings. Most line
profiles show a triangular shape.

Kurtz et al. (\cite{KEM05a}) report a detection of weak RV
oscillations in the Fe\i\,5434~\AA\ line. Our measurements confirm
this result and also provide the following pulsation RV amplitudes
for light and iron-peak elements: 30--50~\ms\ for Na, Ca, Cr, Fe
lines; $\approx$100~\ms\ for Ba\ii\ lines; 150--200~\ms\ for Ti\ii\
lines.



\subsection{HD~9289}
\label{HD9289}

\begin{figure}[!th]
\centering
\firrps{82mm}{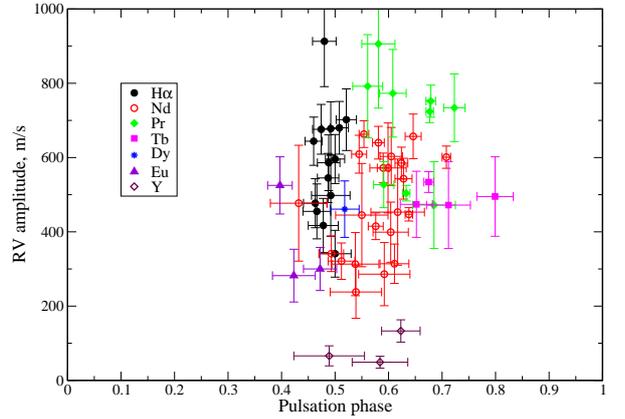}
\caption{The same as in Fig.~\ref{alcir_diagr} but for HD~9289.}
\label{hd9289-rv}
\end{figure}

The amplitude-phase diagrams for the centre-of-gravity measurements in the spectrum of HD~9289 are shown in Fig.~\ref{hd9289-rv}.
Because of the low S/N and significant rotation compared to other stars, we did not perform bisector measurements. The amplitude-phase
diagrams for this star are similar to those for HD~12932.

\subsection{HD~137949 (33 Lib)}
\label{33Lib}

\begin{figure*}[!th]
\centering \firps{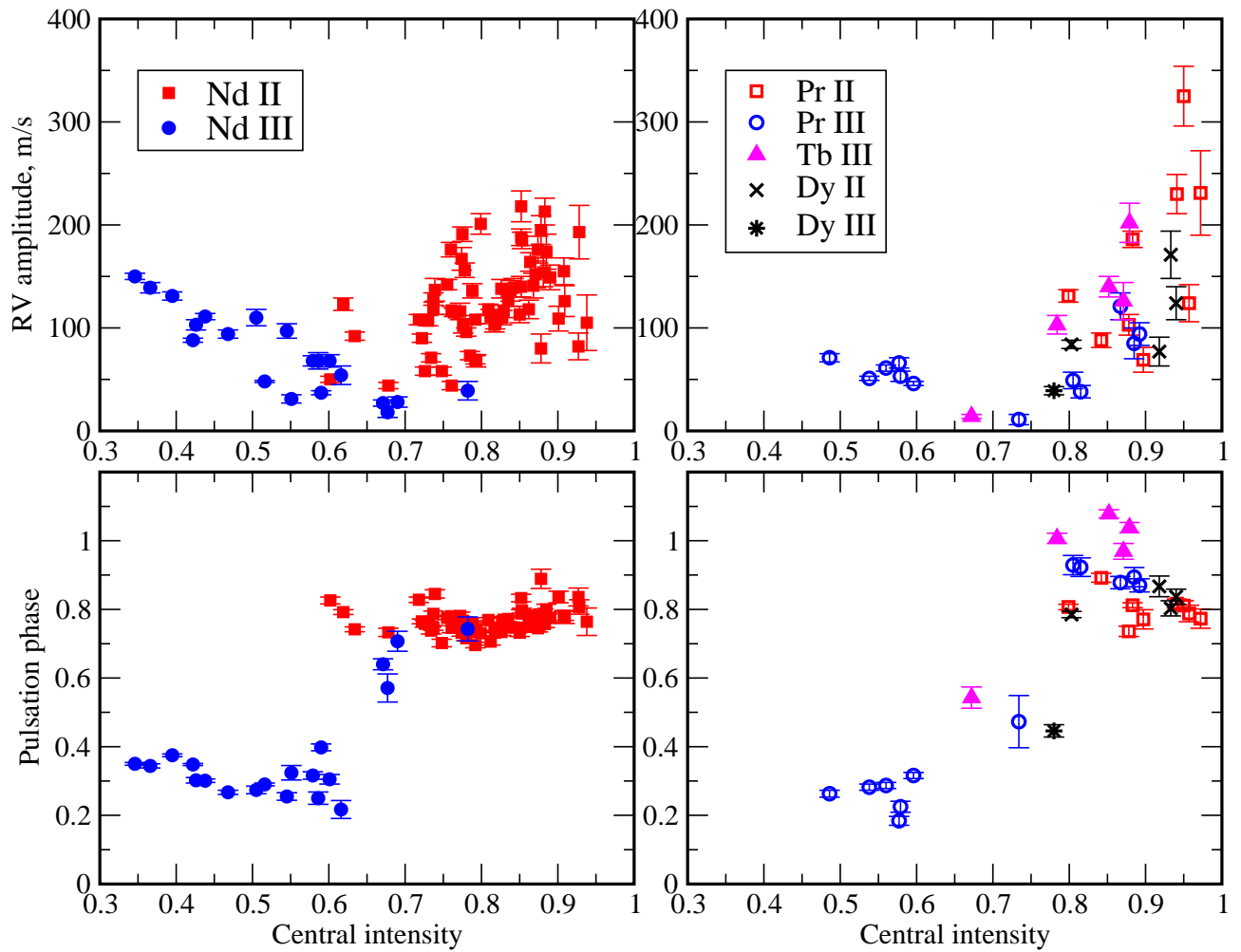} \caption{Amplitude (top)
and phase (bottom) dependence on the central line intensity in the
atmosphere of 33~Lib for the main frequency. Left panels represent
measurements in Nd\ii-\iii\ lines, while right panels show the same
for other elements with lines in the second and third ionisation
states.} \label{33Lib-int}
\end{figure*}

33~Lib shows the most complex pulsational behaviour among all roAp
stars. Mkrtichian et al. (\cite{MHK03}) find nearly anti-phase
pulsations of Nd\ii\ and Nd\iii\ lines, which they attribute to the
presence of pulsation node high in the atmosphere of 33~Lib. This
was confirmed by Kurtz at al. (\cite{KEM05b}), who also find that in
some REE lines the main frequency, corresponding to 8.27 min, and
its harmonic have almost equal RV amplitudes. The high accuracy of
the present observations allows us to study pulsational
characteristics of the 33~Lib atmosphere in more detail. Figure
~\ref{33Lib-int} shows RV amplitudes (top) and phases (bottom)
measured for the main period as a function of the central intensity
of spectral lines. The NLTE Nd stratification analysis by Mashonkina
et al. (\cite{MRR05}) for HD~24712 shows that central intensities of
both Nd\ii\ and Nd\iii\ lines are roughly proportional to there
centre-of-gravity depth formation. Taking into account that the REE
anomaly on which Nd stratification was based is the common feature
for all roAp stars (Ryabchikova et al. \cite{RSMK01}, \cite{RNW04}),
we can expect similar REE gradients in the atmosphere of 33 Lib,
hence, the similar dependence of the central intensities on the
optical depth.

The phase jump is defined very well by Nd lines and is also
indicated by the measurements of Pr, Tb, Dy lines. Our results show
that the phase jump (radial node) does not separate the formation
regions of the lines in consecutive ionisation states as is claimed
by Kurtz at al. (\cite{KEM05b}). Nd\iii, Pr\iii, and Th\iii\ lines
are observed from both sides of the jump. Moreover, a position of
the phase jump relative to line depth provides some evidence that
abundance distributions for at least Pr, Nd, Tb, Dy are similar,
hence one may expect abundance jumps at approximately the same
optical depths in 33~Lib atmosphere.

\begin{figure*}[!th]
\centering
\firps{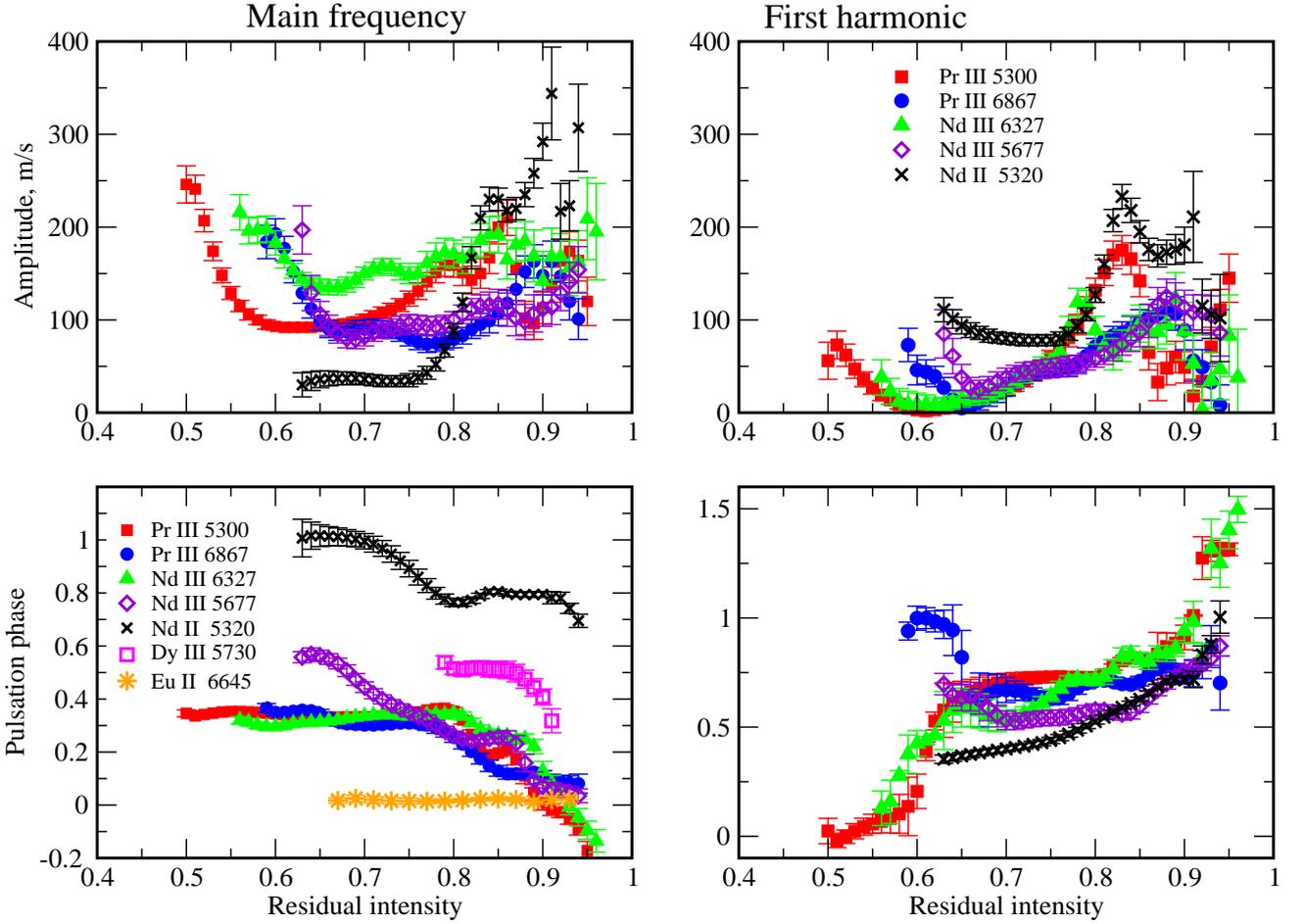}
\caption{Amplitude-phase diagrams for the pulsation bisector measurements in 33~Lib. Amplitudes and phases are shown
for the main period 8.27 min (left panels) and its first harmonic (right panels).}
\label{33Lib-harm}
\end{figure*}

Bisector measurements support the results obtained for the main
frequency from centre-of-gravity measurements and allow us to carry
out a more detailed study of the pulsation properties in 33~Lib.
Figure~\ref{33Lib-harm} shows bisector amplitudes (top) and phases
(bottom) measured in a few representative spectral lines for the
main period 8.27 min (left panels) and its harmonic (right panels).

\begin{figure}[!th]
\centering
\firrps{82mm}{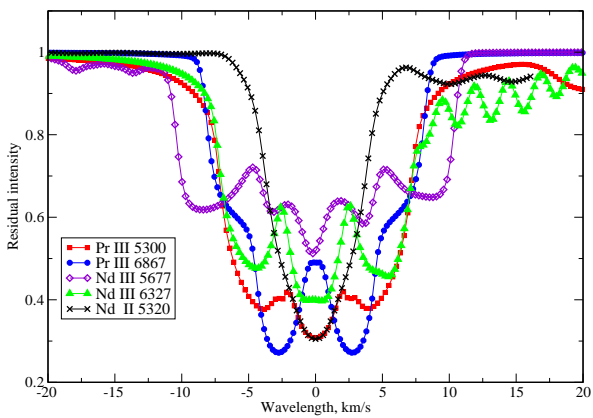}
\caption{Zeeman structure for a set of REE spectral lines in 33~Lib.}
\label{33Lib-prof}
\end{figure}

\begin{figure*}[!v]
\centering
\firrps{82mm}{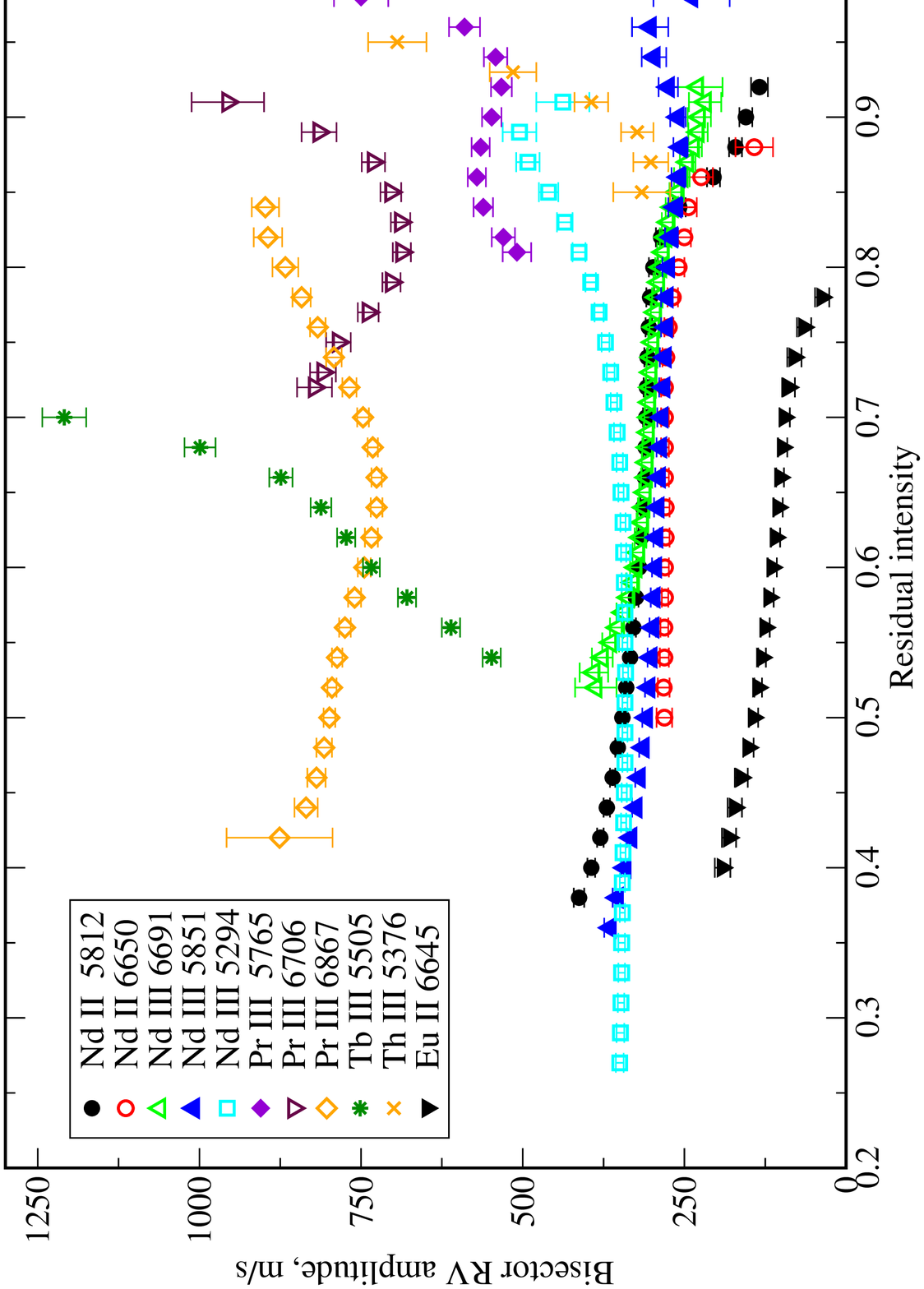}\hspace{0.5cm}\firrps{82mm}{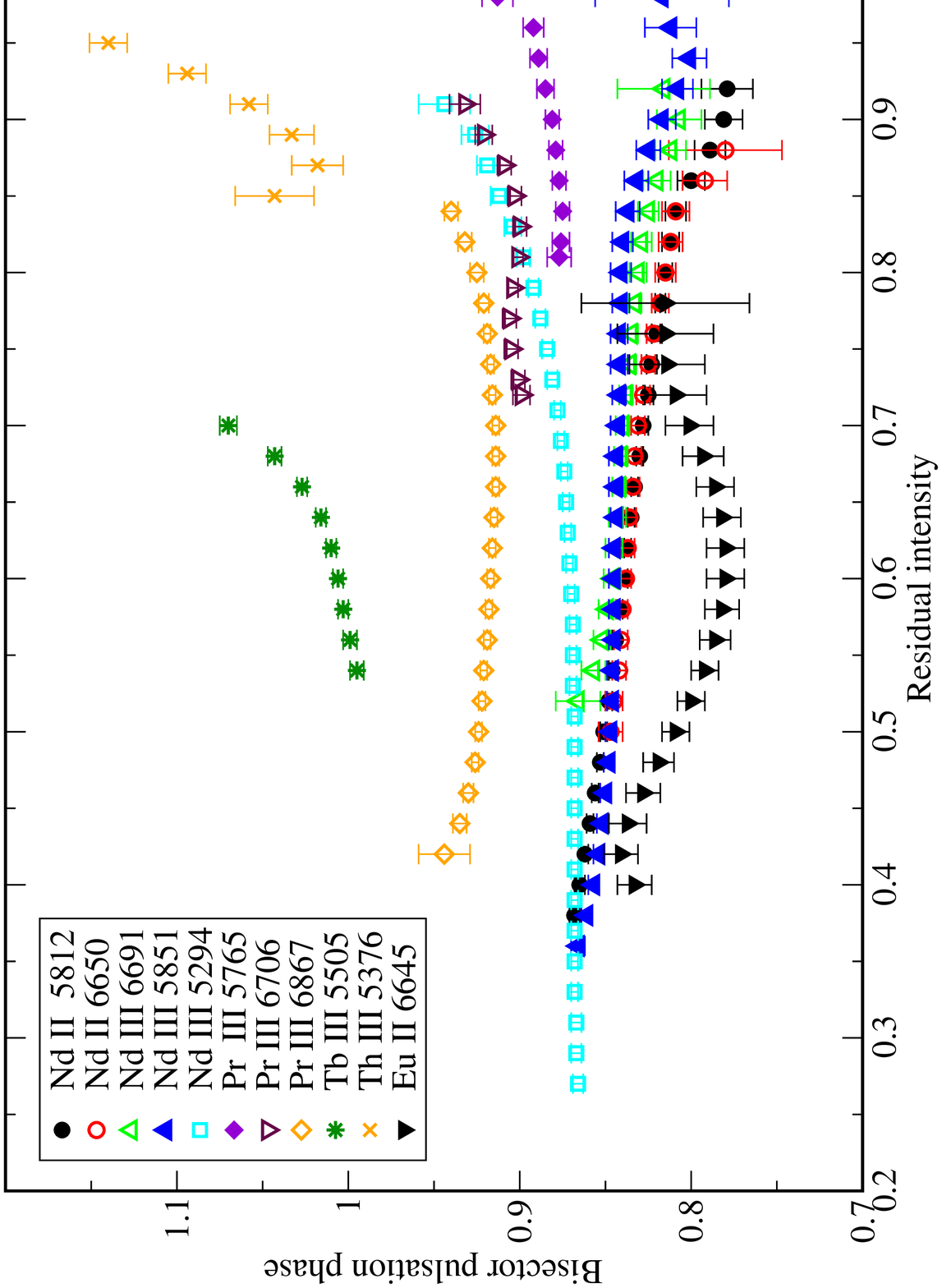}
\caption{Bisector measurements as a function of the line residual intensity in HD~101065. Left panel represents RV amplitudes, right
panel shows phases calculated with the main period from Table~\ref{tbl2}).}
\label{HD101065_prof}
\end{figure*}

\begin{figure*}[!h]
\centering
\firrps{82mm}{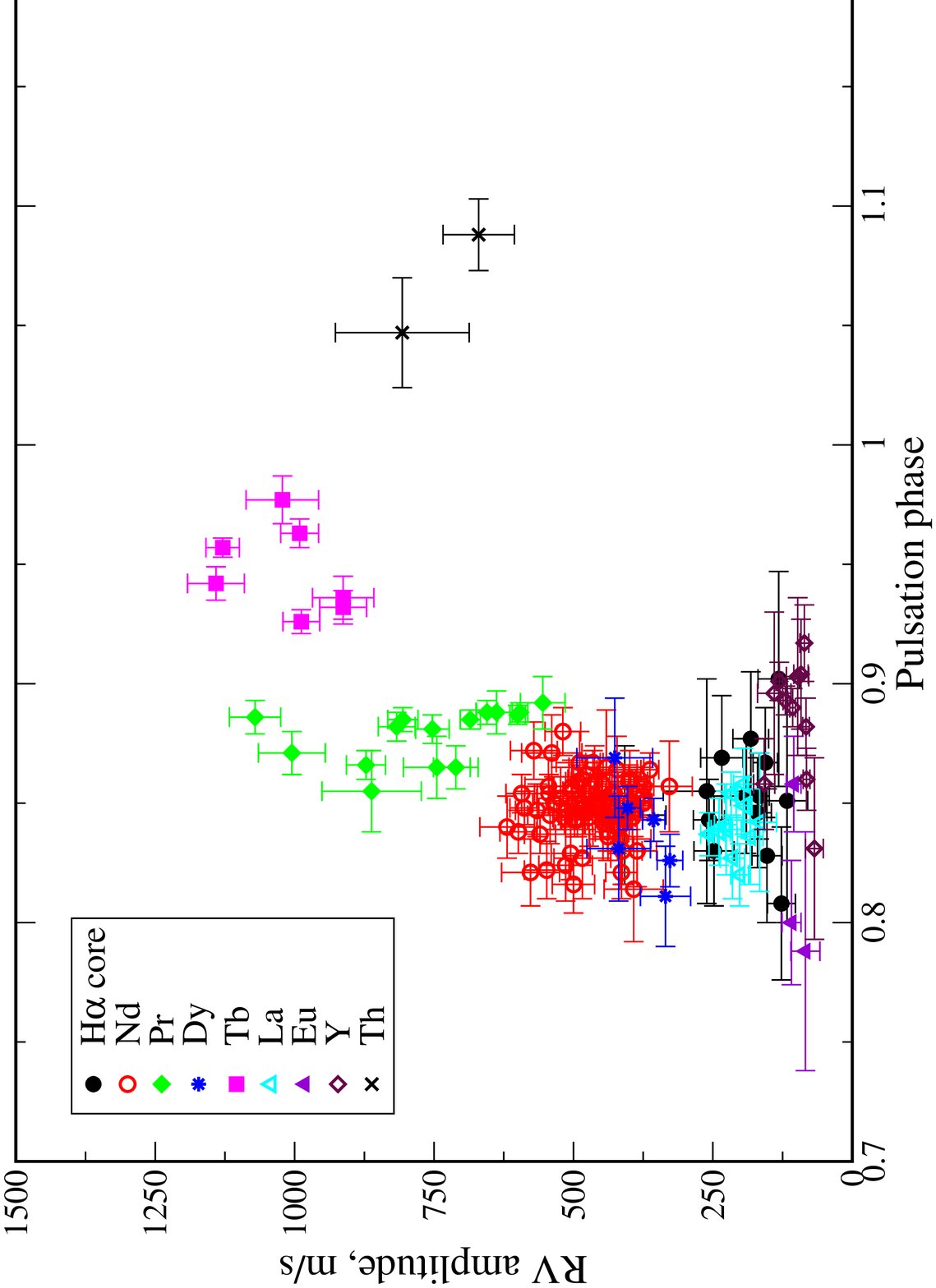}\hspace{0.5cm}\firrps{82mm}{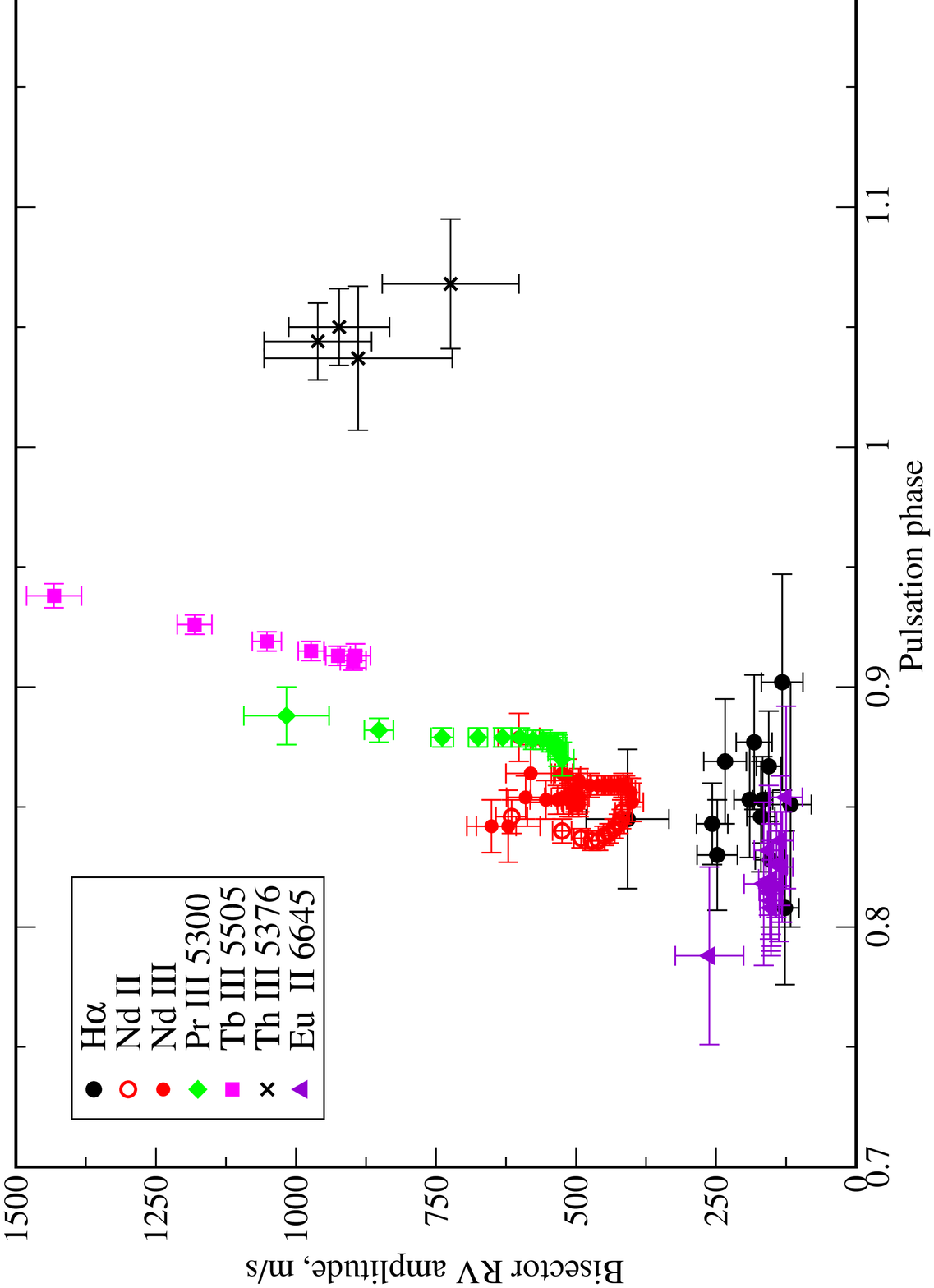}
\caption{The same as in Fig.~\ref{HD101065} but for HD~122970.}
\label{HD122970}
\end{figure*}

\begin{figure*}[!h]
\centering
\firrps{82mm}{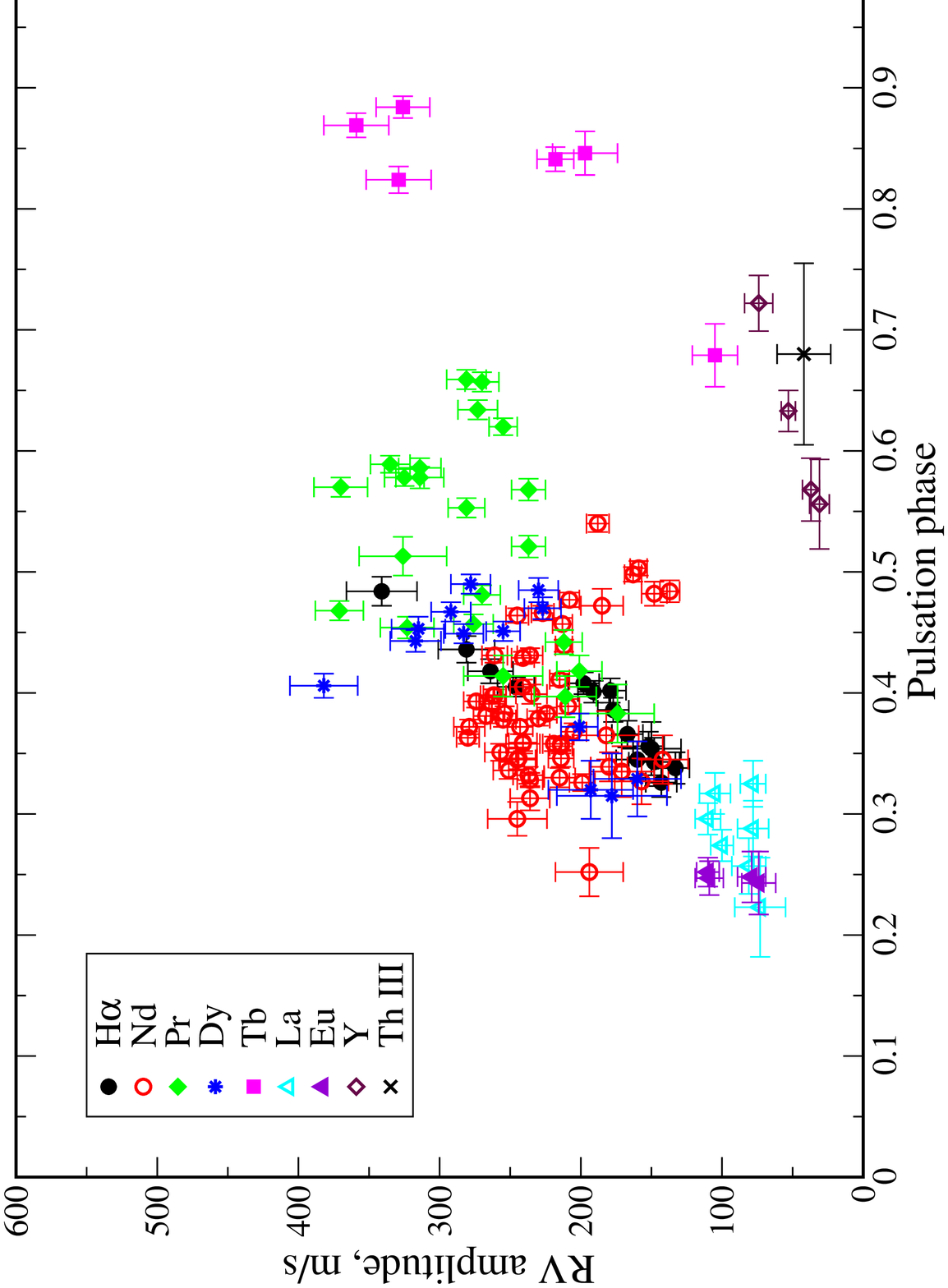}\hspace{0.5cm}\firrps{82mm}{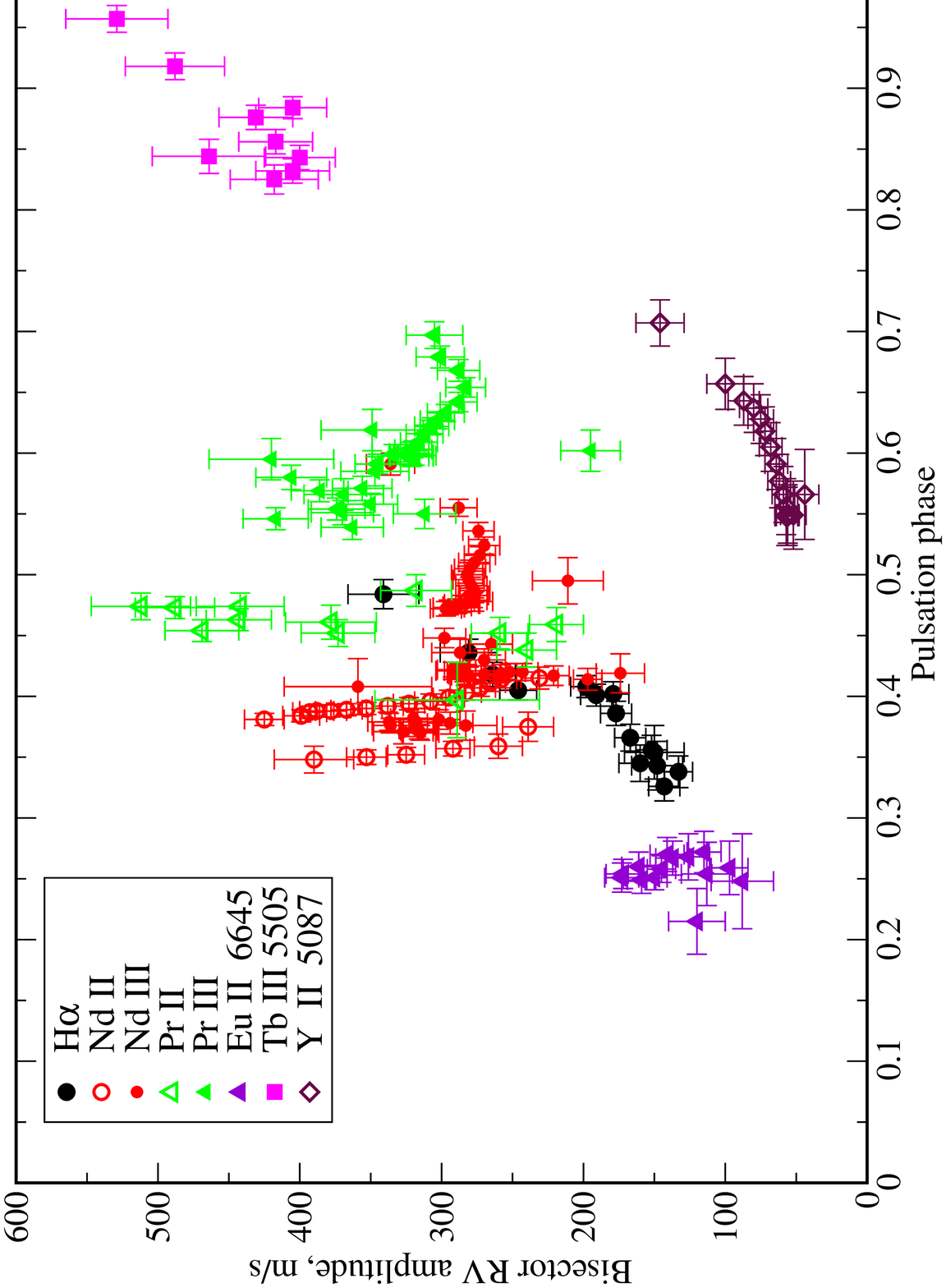}
\caption{The same as in Fig.~\ref{HD101065} but for HD~24712.}
\label{HD24712}
\end{figure*}

\begin{figure}[!h]
\centering
\firrps{82mm}{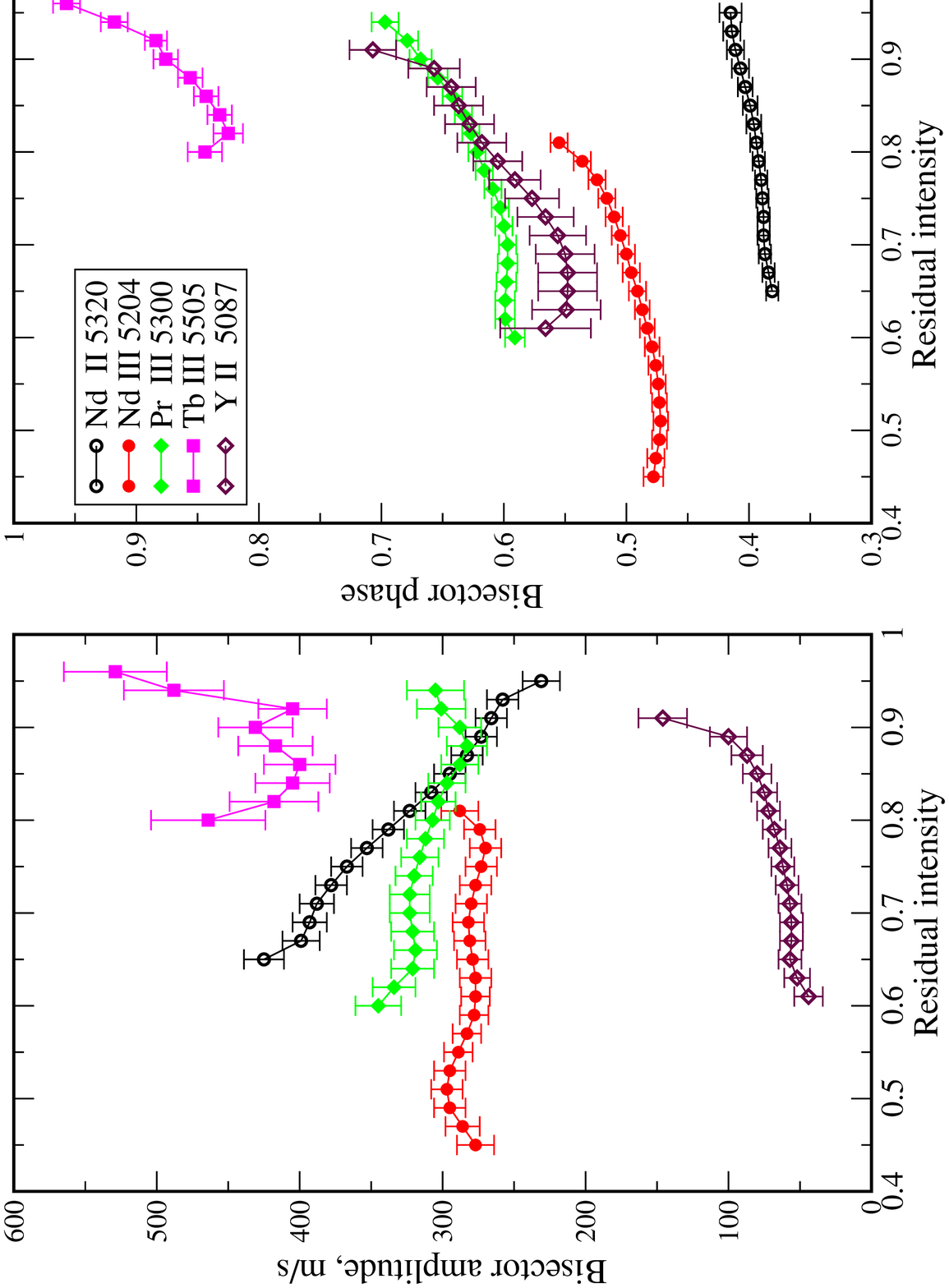}
\caption{The same as in Fig.~\ref{HD122970_int} but for HD~24712.}
\label{HD24712_int}
\end{figure}

\begin{figure*}[!h]
\centering
\firrps{82mm}{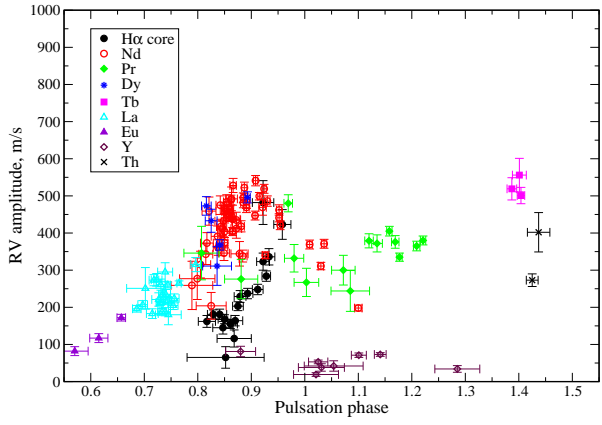}\hspace{0.5cm}\firrps{82mm}{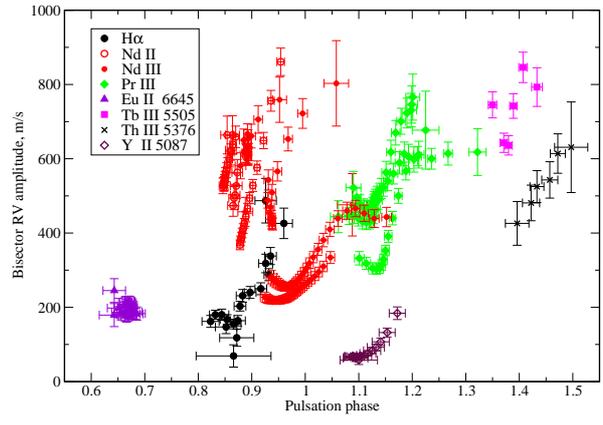}
\caption{The same as in Fig.~\ref{HD101065} but for HD~134214.}
\label{HD134214}
\end{figure*}

\begin{figure*}[!h]
\centering
\firrps{82mm}{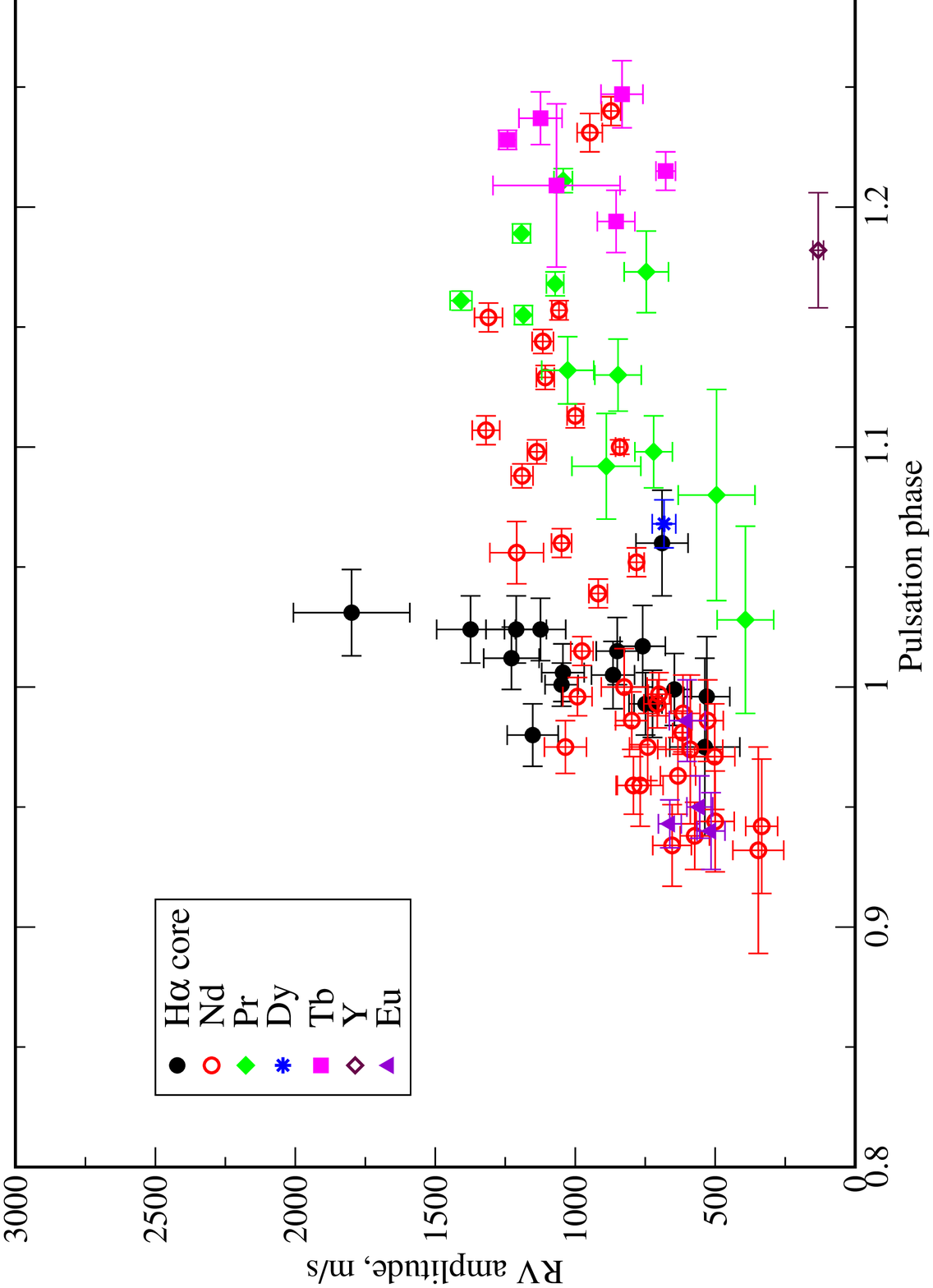}\hspace{0.5cm}\firrps{82mm}{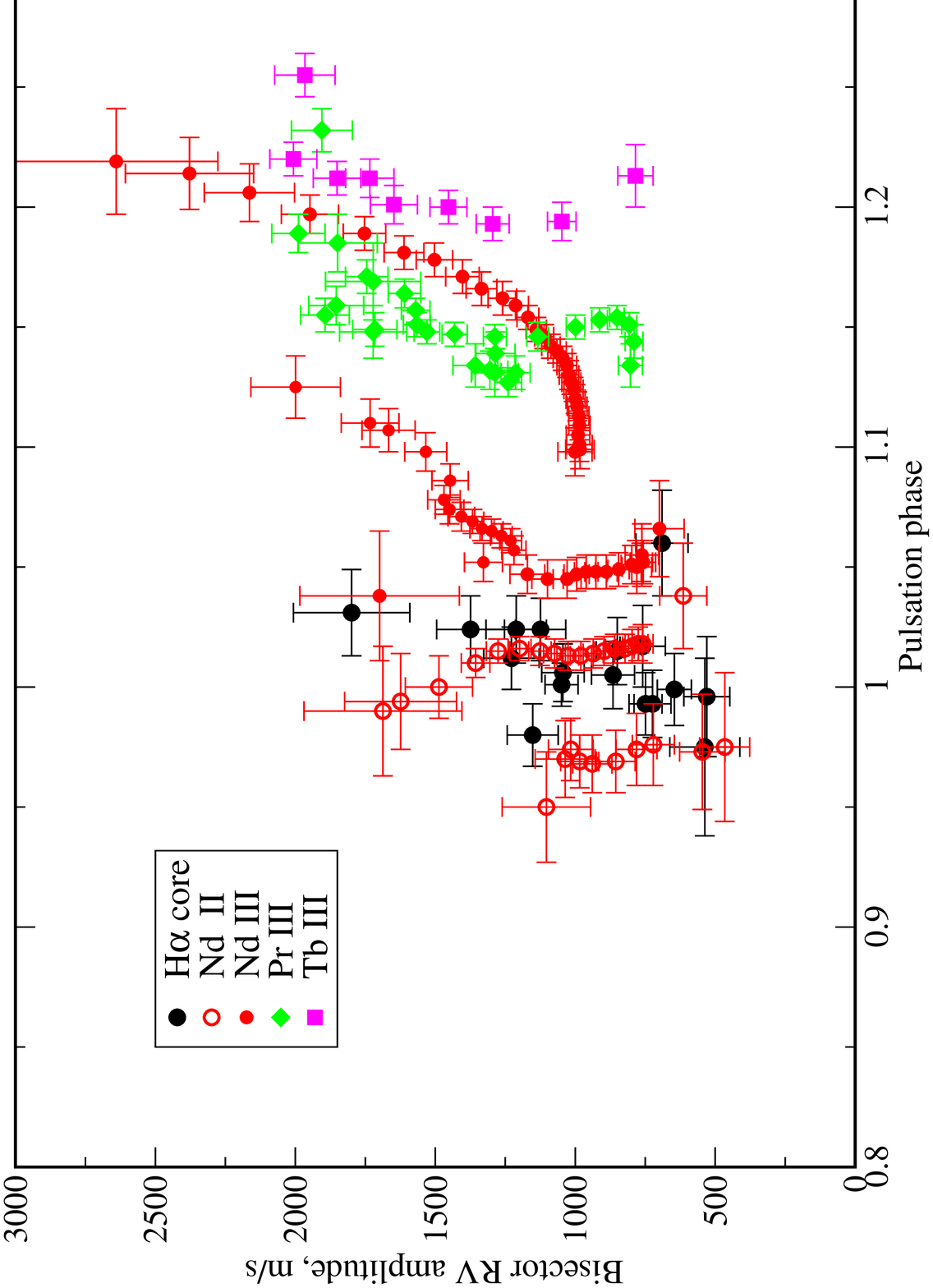}
\caption{The same as in Fig.~\ref{HD101065} but for HD~12932.}
\label{hd12932}
\end{figure*}

\begin{figure}[!h]
\centering
\firrps{82mm}{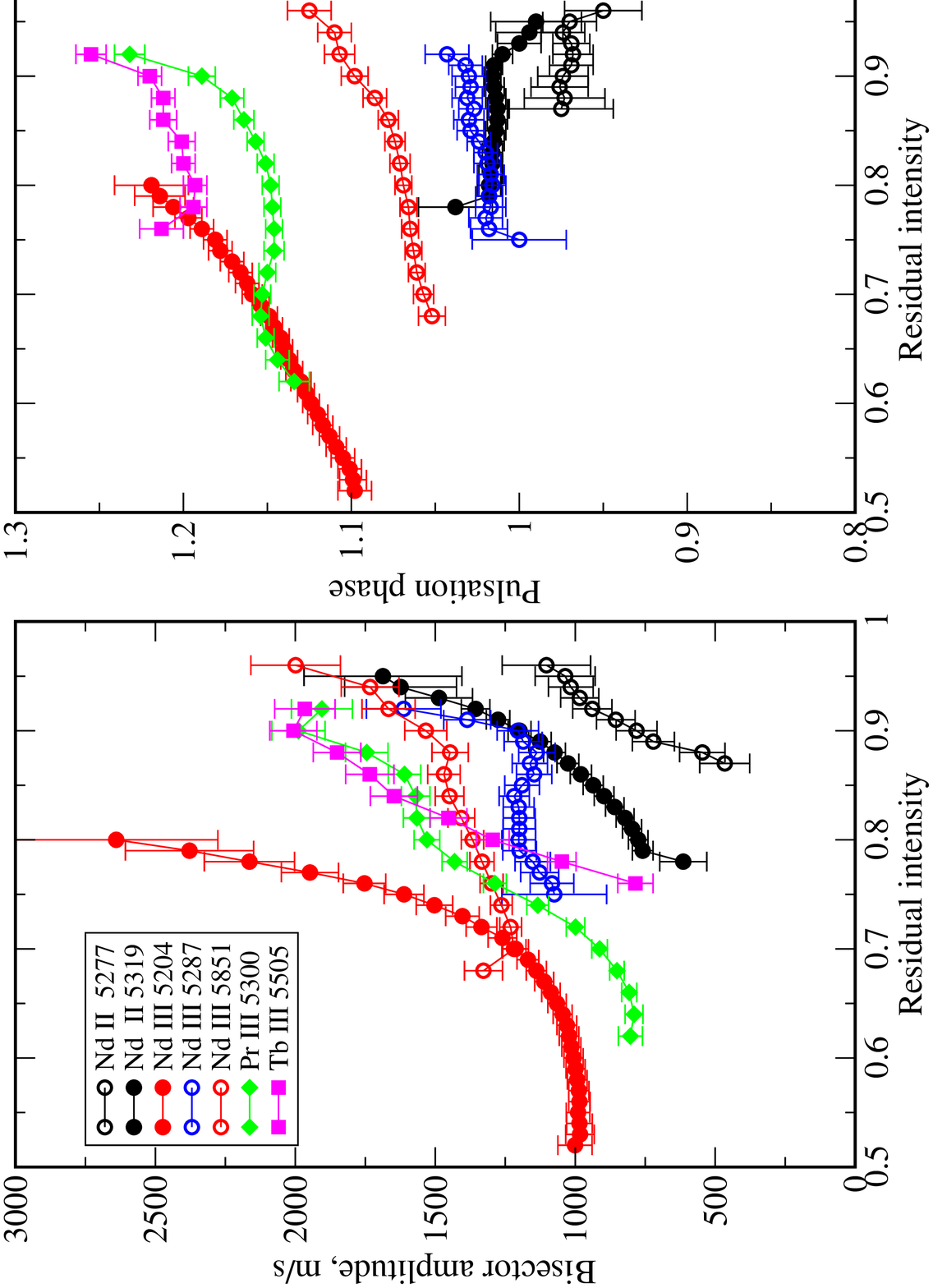}
\caption{The same as in Fig.~\ref{HD122970_int} but for HD~12932.}
\label{HD12932_int}
\end{figure}

\begin{figure*}[!h]
\centering
\firrps{82mm}{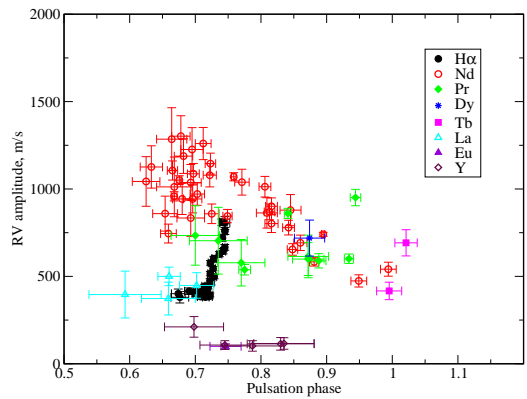}\hspace{0.5cm}\firrps{82mm}{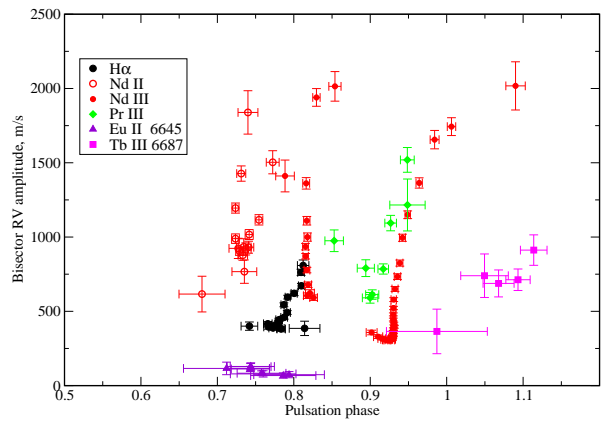}
\caption{The same as in Fig.~\ref{HD101065} but for $\gamma$~Equ.}
\label{gequ-rv}
\end{figure*}

\begin{figure}[!h]
\centering
\firrps{82mm}{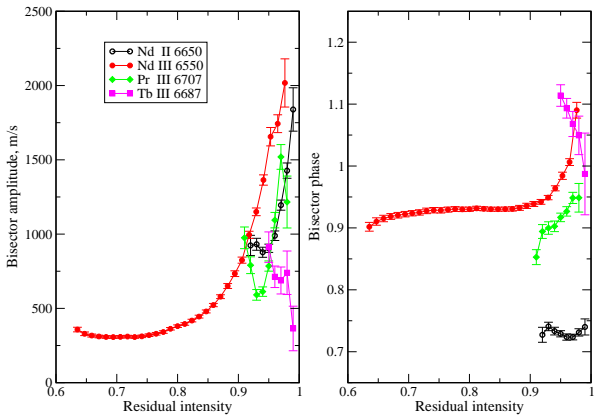}
\caption{The same as in Fig.~\ref{HD122970_int} but for HD~201601.}
\label{HD201601_int}
\end{figure}

\begin{figure}[!h]
\centering
\firrps{82mm}{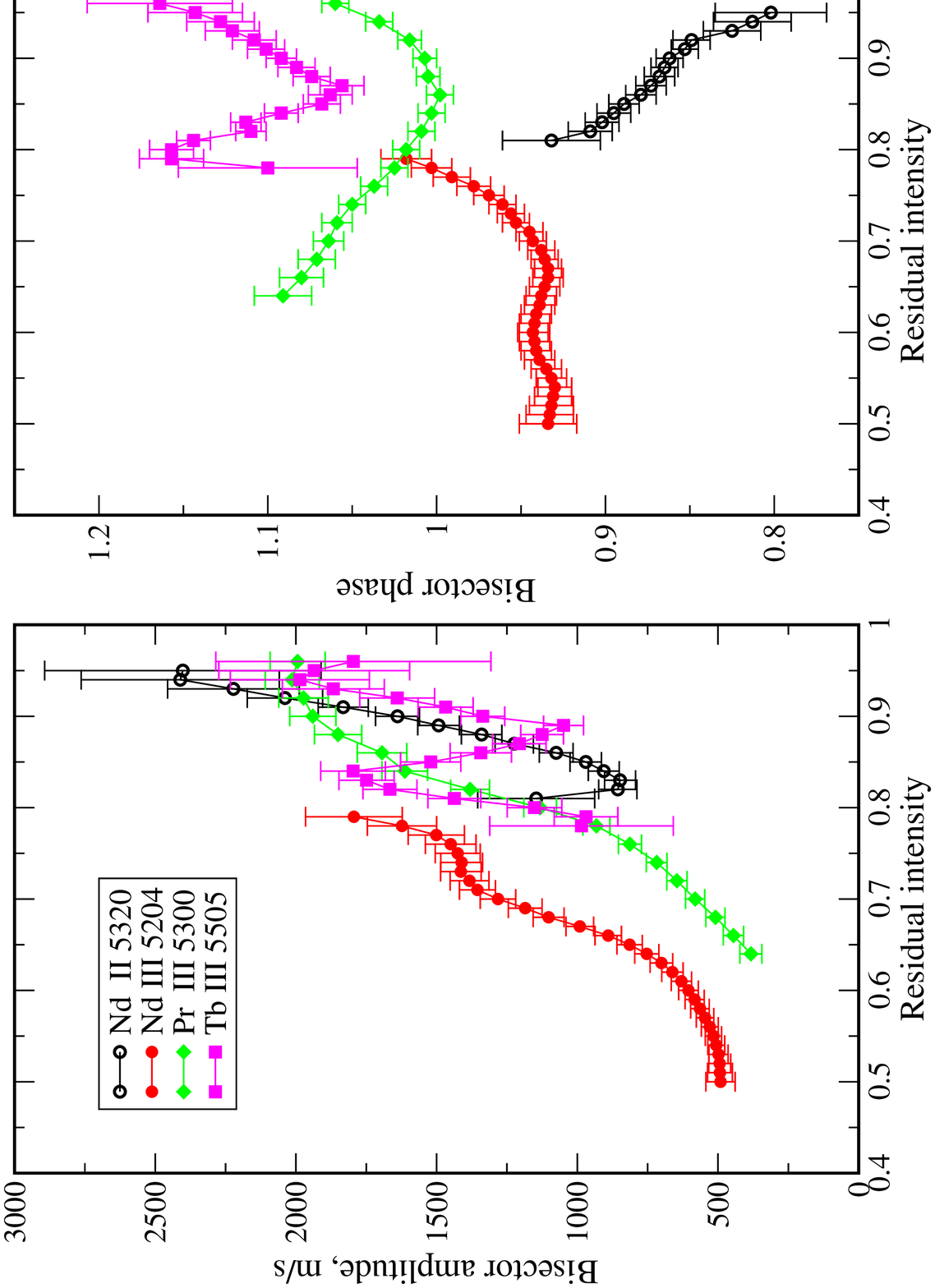}
\caption{The same as in Fig.~\ref{HD122970_int} but for HD~19918.}
\label{HD19918_int}
\end{figure}

\begin{figure*}[!h]
\centering
\firrps{82mm}{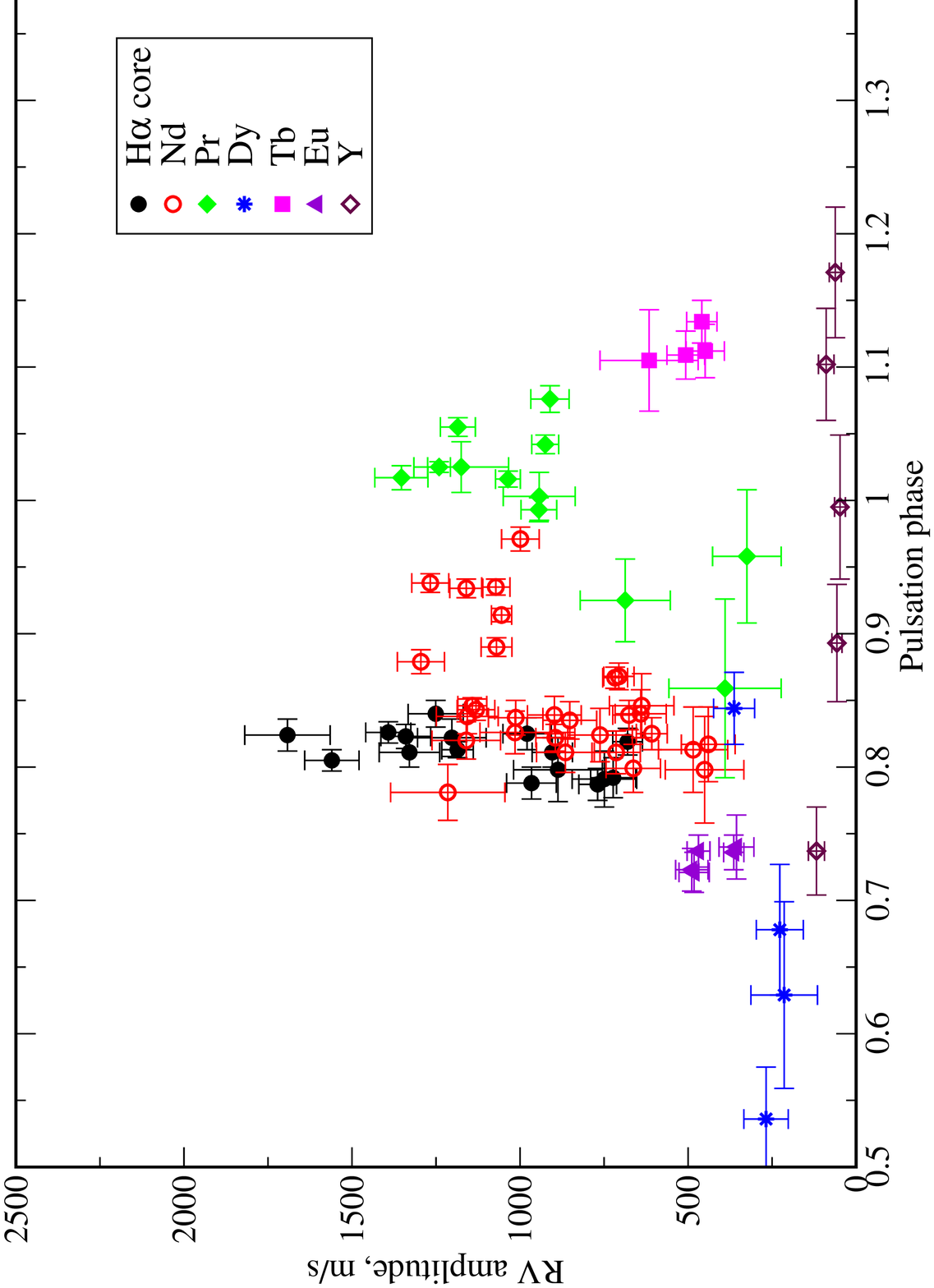}\hspace{0.5cm}\firrps{82mm}{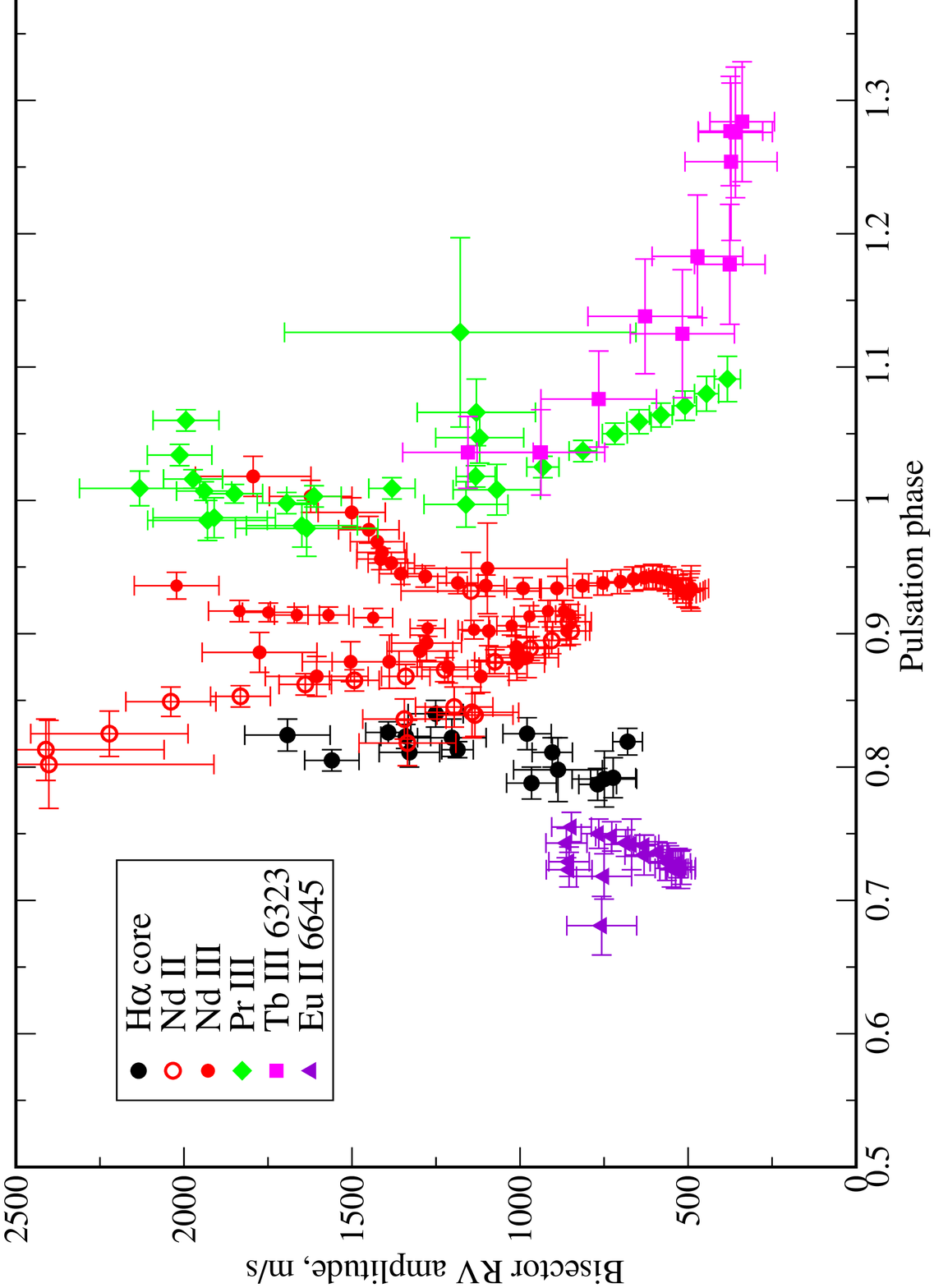}
\caption{The same as in Fig.~\ref{HD101065} but for HD~19918.}
\label{hd19918-rv}
\end{figure*}

We have to emphasise here a major difficulty in the bisector
measurements for the lines splitted in strong magnetic field. Figure
~\ref{33Lib-prof} shows intrinsic theoretical profiles of the lines
from Fig.~\ref{33Lib-int}, calculated with the abundances that
reproduce the observed line intensity. Only the Nd\ii\ 5320~\AA\
line does not exhibit resolved Zeeman components in the intrinsic
profile, therefore this line may provide the `purest' pulsation
distribution across the line profile. Unfortunately, no strong
Nd\iii\ or Pr\iii\ line has negligible Zeeman splitting. The best
and strongest Pr\iii\ 5300~\AA\ and Nd\iii\ 5294~\AA\ lines both
have significant splitting and rather different Zeeman patterns.
Moreover, the Nd\iii\ 5294~\AA\ line is blended in both wings, and
these blends may distort the observed pulsation effects. In
particular, it concerns the RV amplitudes and, to a lesser extent,
pulsation phases. Therefore we chose another Nd\iii\ 6327~\AA\ line
as a representative of Nd\iii\ strong lines. Note that spectral
lines with similar Zeeman patterns and similar intensities (Nd\iii\
5677, 5802, 5851~\AA) have the same bisector velocity amplitude and
phase distributions. Pulsation phases are still similar for the
lines with similar Zeeman patterns but with different intensities
(Nd\iii\ 6327~\AA\ and Pr\iii\ 5300~\AA). Therefore, here we discuss
mainly pulsation phase distribution in the atmosphere of 33~Lib.

Across the profiles of all strong Pr\iii\ and Nd\iii\ lines and the
Nd\ii~5320~\AA\ line, one phase jump is observed for the main period
and two phase jumps are detected for the first harmonic. Kurtz et
al. (\cite{KEM05b}) find only one phase jump for the first harmonic
using bisector measurements of the Nd\iii\,6145~\AA\ line. The best
pulsation distribution is defined by the bisector measurements in
the Pr\iii\,5300~\AA\ line, where it is clearly seen that the
position of the phase jumps corresponds to nearly zero RV
amplitudes, indicating two radial nodes.

Due to the diversity of the pulsational characteristics, it is
difficult to compare pulsation phases for the lines of different
species. However, a phase distribution for the main period
(Fig.~\ref{33Lib-int}, bottom left panel) suggests the existence of
the $\pi$-radian jump between Nd\ii\ and Nd\iii\ lines, which was
found earlier by Mkrtichian et al. (\cite{MHK03}), and another
$\pi$-radian jump between the Eu\ii\ 6645~\AA\ and Dy\iii\ 5730~\AA\
lines. Comprehensive analysis of the variability of many REE lines
in 33~Lib allowed us to obtain a refined picture of the pulsation
node in high atmospheric layers. In particular, we show (see
Fig.~\ref{33Lib-int}, upper panel) that Nd\ii\ and Nd\iii\ lines do
not simply show a $\pi$-radian phase difference, but a continuous
phase trend exists in the RV curves of Nd\iii\ lines, with some of
the doubly ionised Nd lines (e.g., 5286 and 6690~\AA) pulsating with
phases typical of Nd\ii. This appears to be the first demonstration
of the existence of a pulsation node within the formation region of
the lines belonging to the same REE ion. Another remarkable
observation that we made is that harmonic RV variations are the
strongest in the lines that form close to the position of the node.
Thus, the mechanism exciting the first harmonic is probably directly
related to the presence of a radial node in the upper atmosphere of
33~Lib.

The phases of the \ha\ core lie between the phases of the Eu\ii\
line and the wing of the Dy\iii\ 5730~\AA\ line. Interestingly, the
very core of \ha\ has the highest RV amplitude, 500~\ms, and the
last measured point in the Dy\iii\ line wing has very high RV
amplitude, too, although this measurement is not accurate enough.
Nevertheless, this gives us a hint of the line formation depths in
the atmosphere of 33~Lib. Without detailed analysis of the chemical
structure of the atmosphere, it is difficult to suggest a
pulsational scenario for the star, but it is evident that we observe
several waves over the whole atmosphere.

The phase distribution across the Pr\iii\ and Nd\iii\ spectral lines
allows us to estimate the pulsation wave speed. Assuming that we
have one full harmonic wave over the Nd\iii\  and Pr\iii\ lines and
considering the geometric size of the line-formation region in the
stratified case (Mashonkina et al. \cite{MRR05}; Ryabchikova et al.
\cite{RMR06}), we estimated the pulsation wave speed as less than
6~\kms\ for the harmonic period 4.136 min. It is lower than the
sound speed in the 33~Lib atmosphere in adiabatic approximation and
is similar to the pulsation wave speed in another roAp star,
HD~24712 (Ryabchikova et al. \cite{RSW06}). The magnetic field,
which is stronger in 33~Lib than in any other programme star,  may
be a reason for the difference in the observed distribution of the
pulsational characteristics over the atmospheres of 33~Lib and
HD~24712.

What conclusions can be made from these results? Calculations of
linear and non-linear radial pulsations (for example, Fadeev \&
Fokin \cite{FF85}) in adiabatic approximation show that the wave
speed may differ from the sound speed because of the reflection
conditions and of the finite amplitude. Therefore, our results
cannot give any proof of the non-adiabaticity of the pulsations in
roAp stars. However, that the pulsation wave speed is close to the
sound speed strongly supports an acoustic nature for the pulsations
and an absence of the shock waves.

While no extra broadening is required to fit the Fe-peak lines,
3--4~\kms\, is necessary for Y\ii, Eu\ii, Gd\ii, and Er\ii\ lines.
For Nd\ii\ and weak Nd\iii\ and for Pr\iii\ lines, one needs
6--8~\kms, and, finally, for the strong Nd\iii\ and Pr\iii\ lines we
have to introduce \vmacro\,=\,10--12~\kms\ to fit the observed
profiles.

Rather high amplitude pulsations were detected in both Th\iii\ lines
with the phases corresponding to those for Tb\iii\ lines. The RV
amplitudes of the main frequency derived for the lines of Na, Mg,
Si, Ca, Sc Cr, Ti, Fe, Ni  lie in the 30--50~\ms\ range, while the
amplitude of the harmonic is below the detection limit of
$\sim$10~\ms. The Ba lines also reveal pulsations at the main
frequency, but with higher RV amplitudes (70--100~\ms). The Y\ii\
lines have similar amplitudes, which seem to grow with the line
intensity. The Y\ii\ lines also show detectable harmonic amplitudes.
All these lines show pulsation phases in the range of pulsation
phases for the singly ionised REEs.

\section{Summary and discussion}
\label{disc}

We have analysed spectroscopic pulsational variability of ten roAp
stars in detail.  For each object, several hundred spectral lines
were measured and the resulting time series of radial velocity and
bisector variation of selected lines were interpreted with the help
of the least-square fitting technique. We confirm results of the
previous studies, which suggest that non-radial oscillations are
primarily detectable in the lines of heavy elements, especially in
the doubly ionised REE lines. This phenomenon is attributed to the
extreme chemical stratification present in the atmospheres of cool
Ap stars. Due to the segregation of elements under the influence of
radiative diffusion and due to lack of mixing, a thin layer enriched
in heavy elements is created in the upper atmosphere, where rather
weak roAp non-radial oscillations attain significant amplitudes.

Stellar atmosphere modelling that includes stratification, NLTE, and
magnetic field effects is an extremely complex problem. Up to now
only the first preliminary models of different element distributions
over stellar atmosphere based on the spectroscopic observations were
calculated for a few cool Ap stars. Self-consistent diffusion
calculations (LeBlanc \& Monin \cite{LM04}) provide the next step in
our understanding of the Ap star phenomenon. However, with
the diffusion calculations for 39 elements from He to La, LeBlanc \& Monin's
models do not include REE elements, which show the most outstanding
stratification signatures in roAp atmospheres. The predicted
stratification of La is far from the observed one because a small
number of the ionisation states (two) is taken into account and NLTE
effects are not considered. Empirically derived Nd abundance
distribution shows directly that NLTE effects play a crucial role in
the REE stratification studies (Mashonkina et al. \cite{MRR05}).
Based on the empirical element distribution in the atmospheres of
two roAp stars, HD~24712 and $\gamma$~Equ, we conclude that
pulsation amplitudes and, in particular, phases derived from the
lines of different elements correlate with the optical depth.
Therefore, in the absence of the proper atmospheric models, the
amplitude-phase diagrams proposed in the present paper are a
powerful tool in the study of vertical structure of $p$-modes in the
atmospheres of roAp stars.

Our pulsation analysis of the radial velocity variations
demonstrates similarity in the atmospheric pulsation
characteristics. With the help of amplitude-phase diagram analysis
we find that pulsation waves exhibit either a constant phase and
amplitude changing with height or depth-dependence of both
parameters. We interpret the former as a signature of standing
pulsation wave and the latter as evidence of a travelling (running)
wave in stellar atmosphere. Thus, in general, atmospheric
pulsational fluctuations in roAp stars can be represented by a
superposition of standing and running waves.  According to our
results, pulsation waves in three roAp stars, HD~24712, HD~134214,
and $\alpha$~Cir, with the pulsation frequency close to or below the
acoustic cut-off limit, have the running-wave character from the low
atmospheric heights. In the longer period stars, standing waves are
observed up to some of the atmospheric layers, defined by the
formation depth of the lines of specific elements, while running
waves dominate higher in the atmosphere.

We also find that the change from the standing to running character
of pulsation waves depends on the effective temperature of roAp
star. Hotter stars seem to develop the running wave deeper in the
atmosphere than the cooler stars with the same pulsation periods.

In all stars but 33~Lib, independent of the atmospheric and
pulsation parameters, pulsation measurements reveal waves travelling
through the layers defined by the same sequence of chemical species.
The lowest amplitudes are observed for Eu\ii\ lines, then pulsations
propagate in the layers where H$\alpha$ core, Nd, and Pr lines
originate. Pulsation amplitude reaches maximum around these
atmospheric heights and then decreases outwards in most stars. The
RV extremum of the second REE ions is always observed later in time
relative to the variation of singly ionised REEs. The largest phase
shifts and amplitudes are often detected in Tb\iii\ and Th\iii\
lines.  Pulsational variability in the latter lines is detected here
for the first time. Similarity of the pulsation wave propagation
signatures in the studied roAp stars suggests that layers enriched
in different REE species are arranged in approximately the same
vertical order in all stars. For two stars, 33~Lib and HD\,19918, we
find weak but definite RV variability in the lines of iron-peak
elements, confirming previous results by Mkrtichian et al.
(\cite{MHK03}) and Kurtz et al. (\cite{KEM05b}), respectively.

In all stars, spectral lines with the highest RV amplitudes have an
additional broadening varying from 4 to 10--12~\kms\ in terms of the
macroturbulent velocity required to reproduce the observed line
profile shapes. The running wave characteristics usually appear in
the  lines with \vmacro\,$\ge$\,10~\kms. According to Kochukhov et
al. (\cite{KR07}), these are the lines where the usual symmetric
pulsational line profile variability is transformed into an
asymmetric blue-to-red moving pattern. These authors propose
pulsational variability of the line widths, arising from the
periodic expansion and compression of the turbulent layers in the
upper atmospheres of roAp stars as an explanation for the asymmetric
line profile variability pattern. The present study shows that the
position of the turbulent layer in the roAp-star atmospheres is
defined by the formation depth of Pr\iii\ lines in a cooler part of
the programme stars and  by the formation depth of Nd\iii\ lines in
hotter stars. This turbulent layer, which is probably related to the
REE abundance gradients in the upper atmosphere (Mashonkina et al.
\cite{MRR05}), seems to be the key element in the depth-dependence
of the spectroscopic pulsation characteristics in roAp stars.
Abundance gradients may cause a non-standard temperature structure
in Ap-star atmospheres. This is supported by an atmospheric anomaly
that required to explain the peculiar core-wing transition of the
Balmer lines (Kochukhov et al. \cite{cwaT}). Therefore, the
comprehensive investigation of the roAp-star atmosphere, including
chemical diffusion and magnetohydrodynamic modelling of the
interaction of convection, pulsation and magnetic field, is
necessary to fully understand the full variety of the phenomena
associated with a turbulent REE-rich cloud in cool Ap stars.

One of the sample stars, 33~Lib, differs from all other roAp stars
in that, at a given atmospheric heights, it shows a comparable RV
amplitude of the main  frequency and its first harmonic.
Comprehensive analysis of the REE lines of several elements in both
ionisations stages shows that harmonic oscillation emerges close to
the position of the pulsation node located within the REE-rich high
atmospheric layer. This follows from the observation that the lines
showing large double-wave variation are all located close to the
minimum amplitude and to a $\pi$-radian phase jump of the main
frequency.  Thus, the physical mechanism giving rise to harmonic
spectroscopic variability in 33~Lib must be closely related to the
existence of the radial node.

Based on the observation of the two radial nodes for the harmonic
variation across the profiles of a few strongest  Nd\iii\ and
Pr\iii\ lines, we can estimate a radial wavelength and pulsation
wave speed. The latter is below the adiabatic sound speed in the
atmosphere of 33~Lib. The same result was obtained earlier for
another roAp star, HD~24712, using a different approach (Ryabchikova
et al. \cite{RSW06}). These results support an acoustic nature for
the pulsations and reject the idea of the shock wave proposed by
Shibahashi et al. (\cite{SKK04}) for interpreting of the blue-to-red
pulsation pattern, because this model requires supersonic pulsation
motions.

33~Lib is the only star showing definite and direct evidence of
radial nodes in the atmosphere. But the absence of a measurable
pulsation signal in the lines of the elements with  Z\,$\le$\,38 in
most of our programme stars poses a question about the existence of
another node in the lower photospheric layers. Abundance
stratification analyses of cool Ap stars (Ryabchikova et al.
\cite{RPK02}, \cite{RLK05}; Kochukhov et al. \cite{KTR06}), as well
as the model atmosphere with self-consistent diffusion calculations
(LeBlanc \& Monin \cite{LM04}), show that light and iron-peak
elements have a tendency to concentrate in the deeper atmosphere,
below $\log\tau_{5000}=-1$, and to be strongly depleted in the outer
atmospheric layers. Therefore, practically all Fe-peak lines
observed in the optical spectral region are formed in the layers
$-1.5<\log\tau_{5000}<0.0$. The non-adiabatic theoretical pulsation
model for HD~24712 supports the existence of the nodal region near
the photosphere ($\log\tau_{5000}\sim0$) (see Fig.~3 in  Sachkov et
al. \cite{SR06}). The position of the radial node depends on the
frequency (or the radial order) of the mode. As the frequency
increases, the node shifts outwards (Saio, private communication).
According to this model, in the roAp stars HD~24712, HD~134214, and
$\alpha$~Cir, the node is located close to, or slightly above, the
continuum-forming layers. On the other hand, in the longer-period
stars the node should be located below the photosphere (see Figs.~3,
4 in Saio \& Gautschy \cite{SG04}; Fig.~8 in Saio \cite{S05}). For
some intermediate pulsation frequencies, we should detect pulsation
signal in the region of the iron-peak line formation, but this is
not observed. Perhaps, the pulsation amplitude in the lower
atmosphere is diminished to such an extent that pulsations are
undetectable even in anti-node regions.

A discovery of the low-amplitude pulsations in Y\ii\ lines in phase
with the highest-amplitude Pr\iii\ lines delineates another problem.
In the three stars with the shortest pulsation periods, HD~24712,
HD~134214, and $\alpha$~Cir, the Y\ii\ lines produce a secondary
minimum in the amplitude-phase diagrams. These lines have the lowest
detected RV amplitudes and are out of phase by $\pi$-radian with
some other low-amplitude lines, for instance, Eu\ii. Does this
indicate a radial node near the Eu\ii\ line formation heights or do
we see a double-wave structure in the amplitude-phase diagram
depending on the chemical structure of stellar atmospheres?

The amplitude-phase diagrams derived for as many elements and
spectral lines as  possible are proven to be an extremely powerful
tool for investigating the pulsation properties of roAp atmospheres.
Detailed inferences about the vertical mode structure obtained in
our study call for in-depth theoretical  investigation of the
propagation of pulsation waves in magnetically dominated and
chemically stratified  atmospheres. The most important and difficult
theoretical challenge is to recognise and model the physical
processes that are responsible for remarkable observation of
pulsation wave transformation in the REE-rich layer. Why does the
dominant character of pulsational perturbation changes from standing
to running wave? What causes pulsation amplitude to diminish about
certain atmospheric height? Why does isotropic turbulence increase
dramatically in this layer? And finally, what is the origin of the
complex bisector behaviour observed in several roAp stars?
Substantial theoretical developments are needed to resolve these
issues.

\begin{acknowledgements}

We are thankful to A. Fokin and H. Saio for very useful discussions
on modelling pulsations in stellar atmospheres. Resources provided
by the electronic databases (VALD, SIMBAD, NASA's ADS) are
acknowledged. This work was supported by the research grants from
RFBI (04-02-16788a, 06-02-16110a), from the RAS Presidium (Program
``Origin and Evolution of Stars and Galaxies''), from the Swedish
\textit{Kungliga Fysiografiska S\"allskapet} and \textit{Royal
Academy of Sciences} (grant No. 11630102), and  from Austrian
Science Fund (FWF-P17580).

\end{acknowledgements}


\Online

\pagebreak

\begin{scriptsize}
\longtab{3}{
\begin{longtable}{l|r|r|r|r|r|r|r|r|r|r|r}
\caption{\label{tbl3} Pulsation amplitudes (first line) and phases
(second line) calculated with the period highlighted with italics in
Table~\ref{tbl2}. Errors in the last digits are given in
paranthesis. When no pulsation signal was detected, we give the
formal
amplitude solution for the fixed period without phase information}\\
\hline\hline
Wavelength &  \multicolumn{10}{c|}{Amplitude (\ms) / Phase (period fraction)}\\
\cline{2-11}
      ( \AA ) &   HD 24712 & HD 101065  & HD 122970  & HD 134214  & HD 12932   &   HD 19918  &   HD 201601 &   HD 128898 &    HD 9289  & HD 137949\\
\hline
\endfirsthead
\caption{Continued.}\\
\hline\hline
Wavelength &  \multicolumn{10}{c|}{Amplitude (\ms) / Phase (period fraction)}\\
\cline{2-11}
      ( \AA ) &   HD 24712 & HD 101065  & HD 122970  & HD 134214  & HD 12932   &   HD 19918  &   HD 201601 &   HD 128898 &    HD 9289  & HD 137949\\
\hline
\endhead
\hline
\endfoot
\hline
\endfoot
\hline
\multicolumn{11}{l|}{ Fe I - Fe II}\\
\hline
 5405.775 & 7 (~5)    & 7 (~4)    & 14 (~4)   & 11(~4)    &25 (18)    & 37 (15)   &17 (19)    & 9 (~3)    &11(20)     & 17(~2)    \\
          &           &           &           &           &           &0.607 (67) &           &           &           &0.686(21)  \\
 5410.910 &12 (~4)    &           & 11 (~4)   & 6 (~3)    &16 (14)    & 52 (18)   &60 (24)    & 6 (~3)    &23(12)     & 14(~1)    \\
          &           &           &           &           &           &0.650 (56) &           &           &           &0.632(22)  \\
 5415.199 & 8 (~4)    &21 (~7)    & 2(~4)     & 9 (~3)    &18(13)     &21(~9)     &14 (14)    & 5 (~4)    &48(23)     & 13(~1)    \\
          &           &           &           &           &           &0.624(70)  &           &           &           &0.396(24)  \\
 5424.068 & 4 (~4)    &           & 6(~4)     & 2 (~3)    &13(~9)     &29(~8)     &31 (18)    & 2 (~5)    &38(24)     & 25(~2)    \\
\hline
\end{longtable}
}
\end{scriptsize}

\begin{thebibliography}{}

\bibitem[1998]{AKM98}
Audard, N., Kupka, F., Morel, P., Provost, J., \& Weiss, W.W. 1998, \aap, 335, 954
\bibitem[1992]{Babel92}
Babel, J. 1992, \aap, 258, 449
\bibitem[1998]{BB98}
Baldry, I. K., Bedding, T.R., Viskum, M., Kjeldsen, H., \& Frandsen, S. 1998, \mnras, 295, 33
\bibitem[2000]{BB00}
Baldry, I. K., \& Bedding, T. R. 2000, \mnras, 318, 341
\bibitem[2002]{Balona02}
Balona, L. A. 2002, \mnras, 337, 1059
\bibitem[1999]{dream99}
Bi\'emont, E., Palmeri, P., \& Quinet, P. 1999, \apss, 635, 2691.
\bibitem[2005]{B05}
Burki, G. et al. 2005, GENEVA photometric database, Geneva Observatory
\bibitem[2000]{CWA}
Cowley, C. R., Hubrig, S., Ryabchikova, T., et al. 2001, \aap, 367, 939
\bibitem[2000]{CRK00}
Cowley, C. R., Ryabchikova, T., Kupka, F., et al. 2000, \mnras, 317, 299
\bibitem[2003]{Nd2-03}
Den Hartog, E. A., Lawler, J. E., Sneden, C., \& Cowan, J. J.  2003, \apjs, 148, 543 
\bibitem[2006]{Gd2}
Den Hartog, E. A., Lawler, J. E., Sneden, C., \& Cowan, J. J.  2006, \apjs, 167, 292 
\bibitem[2005]{EKM05}
Elkin, V. G., Kurtz, D. W., \& Mathys, G. 2005, \mnras, 364, 864
\bibitem[1985]{FF85}
Fadeev, Yu. A., \& Fokin, A. V. 1985, \apss, 111, 355
\bibitem[1998]{HM98}
Hauck, B., \& Mermilliod, M. 1998, \aaps, 129, 431
\bibitem[1986]{HB86}
Horne, J. H., \& Baliunas, S. L. 1986, \apj, 302, 757
\bibitem[1998]{KH98}
Kanaan, A., \& Hatzes, A. P. 1998, \apj, 503, 848
\bibitem[2004]{K04}
Kochukhov, O. 2004, \apj, 615, L149
\bibitem[2005]{K05}
Kochukhov, O. 2005, \aap, 438, 219
\bibitem[2006a]{synthmag06}
Kochukhov, O. 2006a, in {\it Magnetic Stars 2006}, eds. I.I. Romanyuk and D. O. Kudryavtsev, in press
\bibitem[2006b]{K06}
Kochukhov, O. 2006b,
Comm. in Asteroseismology, 149, in press
\bibitem[2001a]{KR01a}
Kochukhov, O., \& Ryabchikova, T. 2001a, \aap, 374, 615
\bibitem[2001b]{KR01b}
Kochukhov, O., \& Ryabchikova, T. 2001b, \aap, 377, L22
\bibitem[2002]{cwaT}
Kochukhov, O., Bagnulo, S., \& Barklem, P.S. 2002, \apj, 578, L75
\bibitem[2004]{KRP04}
Kochukhov, O., Ryabchikova, T., \& Piskunov, N. 2004, \aap, 415, L13
\bibitem[2006]{KTR06}
Kochukhov, O., Tsymbal, V., Ryabchikova, T., Makaganyk, V., \& Bagnulo, S. 2006, \aap, 460, 831
\bibitem[2007]{KR07}
Kochukhov, O., Ryabchikova, T., Weiss, W. W., Landstreet, J. D., \& Lyashko, D. 2007, \mnras, in press (astro-ph/0612761)
\bibitem[1997]{KNK97}
K\"unzli, M., North, P., Kurucz, R. L., \& Nicolet, B. 1997, \aaps, 122, 51
\bibitem[1994]{acir}
Kurtz, D. W., Sullivan, D. J., Martinez, P., \& Tripe, P. 1994, \mnras, 270, 674
\bibitem[2000]{KM00}
Kurtz, D. W., \& Martinez, P. 2000, Baltic Astron., 9, 253
\bibitem[2005a]{KEM05a}
Kurtz, D. W., Elkin, V. G., \& Mathys, G. 2005a, in {\it Element Stratification in Stars: 40 Years
of Atomic Diffusion}, eds. G. Alecian, O. Richard \& S. Vauclair, EAS Publ. Ser., 17, 91
\bibitem[2005b]{KEM05b}
Kurtz, D. W., Elkin, V. G., \& Mathys, G. 2005b, \mnras, 358, L6
\bibitem[2006]{KEM06}
Kurtz, D. W., Elkin, V. G., \& Mathys, G. 2006, \mnras, 370, 1274
\bibitem[1996]{KRW96}
Kupka, F., Ryabchikova, T. A., Weiss, W. W., et al. 1996, \aap, 308, 886
\bibitem[1999]{vald299}
Kupka, F., Piskunov, N., Ryabchikova, T. A., Stempels, H. C., \& Weiss, W. W., 1999, \aaps, 138, 119
\bibitem[2001]{La2}
Lawler, J. E., Bonvallet, G., \& Sneden, C. 2001a, \apj, 556, 452  
\bibitem[2006]{Sm2}
Lawler, J. E., Den Hartog, E. A., Sneden, C. \& Cowan, J. J. 2006, \apjs, 162, 227  
\bibitem[2004]{LM04}
LeBlanc F., \& Monin D. 2004, in {\it The A-Star Puzzle}, Proc. IAU Symp. No.224, eds. J. Zverko, W.W. Weiss,
   J. \,\v{Z}i\v{z}\v{n}ovsk\'{y}, S.J. Adelman, Cambridge University Press, p. 193.
\bibitem[1978]{L78}
Lucke, P. B. 1978, \aap, 64, 367
\bibitem[2006]{MDI06}
L\"uftinger, T., Kochukhov, O., Ryabchikova, T., Weiss, W. W., \& Ilyin, I. 2006,
in {\it Magnetic Stars 2006}, eds. I.I. Romanyuk and D. O. Kudryavtsev, in press
\bibitem[2005]{MRR05}
Mashonkina, L., Ryabchikova, T., \& Ryabtsev, A. 2005, \aap, 441, 309
\bibitem[2000]{MHP00}
Mkrtichian, D. E., Hatzes, A. P., \& Panchuk, V. E. 2000, in
{\it Variable Stars as Essential Astrophysical Tools}, eds. C. Ibanoglu, Kluwer Academic
Pulishers, 405
\bibitem[2003]{MHK03}
Mkrtichian, D. E., Hatzes, A. P., \& Kanaan, A. 2003, \mnras, 345, 781
\bibitem[1985]{MD85}
Moon, T. T., \& Dworetsky, M. M. 1985, \mnras, 217, 305
\bibitem[1993]{N93}
Napiwotzki, R/. Sch\"{o}nberner, D., \& Wenske, V. 1993, \aap, 268, 653
\bibitem[1999]{P99}
Piskunov, N. E. 1999, in 2nd International Workshop on
Solar Polarization, eds. K. Nagendra and J. Stenflo, Kluwer Acad. Publ. ASSL, 243, 515
\bibitem[1995]{R95}
Rogers, N. Y. 1995, Comm. in Asteroseismology, 78
\bibitem[1997]{RLG97}
Ryabchikova, T. A., Landstreet, J. D., Gelbmann, M. J., et al. 1997, \aap, 327, 1137
\bibitem[2000]{RSH00}
Ryabchikova, T. A., Savanov, I. S., Hatzes, A. P., Weiss, W. W., \& Handler, G. 2000, \aap, 357, 981
\bibitem[2001]{RSMK01}
Ryabchikova, T. A., Savanov, I. S., Malanushenko, V.P., \& Kudryavtsev, D.O. 2001, Astron. Rep., 45, 382
\bibitem[2002]{RPK02}
Ryabchikova, T., Piskunov, N., Kochukhov, O., et al. 2002, \aap, 384, 545
\bibitem[2003]{RWL03}
Ryabchikova, T., Wade, G., \& LeBlanc, F. 2003, in IAU Symposium No. 210, Modelling of Stellar
Atmospheres, eds. N.E. Piskunov, W.W. Weiss, \& D.F. Gray, ASP p.301
\bibitem[2004]{RNW04}
Ryabchikova, T., Nesvacil, N., Weiss, W. W, Kochukhov, O., \& St\"utz, Ch., 2004, \aap, 423, 705
\bibitem[2005b]{RLK05}
Ryabchikova, T., Leone, F., \& Kochukhov, O. 2005b, \aap, 438, 973
\bibitem[2005a]{RWA05}
Ryabchikova T., Wade, G. A., Auri\`ere, M., et al. 2005a, \aap, 429, L55
\bibitem[2006a]{RRKB06}
Ryabchikova, T., Ryabtsev, A., Kochukhov, O., \& Bagnulo, S. 2006a, \aap, 456, 329
\bibitem[2006b]{RMR06}
Ryabchikova, T., Mashonkina, L., Ryabtsev, A., Kildiyarova, R., \& Khristoforova, M. 2006b, Comm. in Asteroseismology, 149, in press
\bibitem[2007]{RSW06}
Ryabchikova, T., Sachkov, M., Weiss, W. W., et al. 2007, \aap, 462, 1103
\bibitem[2004]{sach04}
Sachkov, M., Ryabchikova, T., Kochukhov, O., et al. 2004, in
{\it Variable Stars in the Local Group}, eds. D.W. Kurtz \& K.R. Pollard, ASP Conf. Ser., 310, 208
\bibitem[2006]{SR06}
Sachkov, M., Ryabchikova, T., Bagnulo., et al. 2006, MemSAI, 77, 397
\bibitem[2005]{S05}
Saio, H. 2005, \mnras, 360, 1022
\bibitem[2004]{SG04}
Saio, H., \& Gautschy, A. 2004, \mnras, 350, 485
\bibitem[1999]{SMR99}
Savanov, I. S., Malanushenko, V. P., \& Ryabchikova, T. R. 1999, Astron. Lett., 25, 802
\bibitem[2004]{SKK04}
Shibahashi H., Kurtz D. W., Kambe E., Gough D. O. 2004, in {\it IAU Symposium No. 224,
The A-star Puzzle}, eds. J.\,Zverko, J.\,\v{Z}i\v{z}\v{n}ovsk\'{y}, \& S.J.\,Adelman,
W.W.\,Weiss, Cambridge University Press, IAUS~224, 829
\bibitem[2003]{tsymbal}
Tsymbal, V., Lyashko, D., \& Weiss, W. W. 2003, in {\it Modelling of Stellar Atmiiospheres},
IAU Symp. No. 210, eds. N. Piskunov, W.W.Weiss, D.F. Gray, ASP, E49
\end{thebibliography}
\end{document}